\documentclass[11pt,a4paper,twocolumn]{article}

\usepackage{float}

\usepackage[affil-it]{authblk}
\usepackage{graphicx}
\usepackage{fullpage,amsfonts,lineno,setspace}
\usepackage{hyperref}

\usepackage{epstopdf}

\usepackage{amssymb,amsmath}

\usepackage[compact]{titlesec}
\titlespacing{\section}{0pt}{0.4cm}{0.1cm}
\titlespacing{\subsection}{0pt}{0.3cm}{0.1cm}
\titlespacing{\subsubsection}{0pt}{0.2cm}{0.1cm}

\newcommand{\beq}{\begin{equation}}
\newcommand{\eeq}{\end{equation}}
\newcommand{\beaq}{\begin{eqnarray}}
\newcommand{\eeaq}{\end{eqnarray}}

  \renewenvironment{thebibliography}[1]{%
    \begin{oldthebibliography}{#1}%
      \setlength{\parskip}{0ex}%
      \setlength{\itemsep}{0ex}%
  }%
  {%
    \end{oldthebibliography}%
  }

\begin{document}

\title{Cities and Regions in Britain through hierarchical percolation}

\author[1]{Elsa Arcaute\thanks{Corresponding author, email: e.arcaute@ucl.ac.uk}}
\author[1]{Carlos Molinero}
\author[2,1]{Erez Hatna}
\author[3,1]{Roberto Murcio}
\author[4,1]{\\Camilo Vargas-Ruiz}
\author[1]{A. Paolo Masucci}
\author[1]{Michael Batty}

\affil[1]{Centre for Advanced Spatial Analysis (CASA), University College London, UK}
\affil[2]{Center for Advanced Modeling, The Johns Hopkins University, USA}
\affil[3]{Consumer Research Data Centre, Geography, University College London, London, UK}
\affil[4]{4 Prospective Labs Ltd., London, UK}


\date{}


\twocolumn[
\begin{@twocolumnfalse}
\maketitle

\vspace{-10mm}


\small{$^*$ Corresponding author, email: e.arcaute@ucl.ac.uk}\\

\vspace{2mm}

{\bf Keywords:} percolation theory, urban hierarchies, city boundaries, fractal dimension, street networks

{\bf Abbreviations:} GB, Great Britain


\begin{abstract} 

Urban systems present hierarchical structures at many different scales.  These are observed as administrative regional delimitations which are the outcome of complex geographical, political and historical processes which leave almost indelible footprints on infrastructure such as the street network. In this work we uncover a set of hierarchies in Britain at different scales using percolation theory on the street network and on its intersections which are the primary points of interaction and urban agglomeration.
At the larger scales, the observed hierarchical structures can be interpreted as regional fractures of Britain, observed in various forms, from natural boundaries, such as National Parks, to regional divisions based on social class and wealth such as the well-known North-South divide. At smaller scales, cities are generated through recursive percolations on each of the emerging regional clusters. We examine the evolution of the morphology of the system as a whole, by measuring the fractal dimension of the clusters at each distance threshold in the percolation. We observe that this reaches a maximum plateau at a specific distance. The clusters defined at this distance threshold are in excellent correspondence with the boundaries of cities recovered from satellite images, and from previous methods using population density.
\vspace*{5mm}

\end{abstract}
\end{@twocolumnfalse}
]


\section{Introduction}

Countries are the historical outcome of the unification and the fragmentation of regions and communities.
Although many of these processes are the result of an imposed organisation devised through administrative boundaries, some others hold communities together through strong ideological ties at the regional level, creating a strong sense of belonging.
These processes are intricate cultural, political and socio-historical pathways, that have left footprints in the way communities emerge, organise, trade and change spatially. 
These footprints are contained in the street patterns and network, which are the main proxies for communication and exchange between settlements. It is hence not surprising that these also encompass the socio-economic history of the region. 
Despite the constant change and renewal of streets, we show in this work, that these footprints can still be identified and recovered.

We focus on Britain, whose road network has been evolving for over 2000 years. 
Its origins can be traced back to the iron age with the Celts, but it was during the Roman occupation that a rapid expansion of the roads took place, and a network was established. 
In the last 200 years, this has been subjected to ongoing urban growth, and to adaptations for new extensions and modes of transport.
Britain presents a rich regional structure, whose tensions lead to a fractured landscape, driven by ideologies and socio-historical trajectories.
One example is the surge for the need to call for a referendum on Scottish independence (2014), and the rise of the Scottish Nationalists in the recent elections of 2014-2015.
Discerning the emergent fractions that are independent of administrative provisions, but highly tied to their trajectories and ideologies, is of considerable relevance for the understanding of the dynamical regional functioning of a country.
In a different paper \cite{MolineroVoting_arXiv2015}, we corroborate the high correlation between regional divisions and polarisation represented through voting patterns.

In this work we investigate whether the regional fractures of Britain can be observed through an underlying hierarchical structure which can be recovered from the road network. 
From the beginnings of locational analysis \cite{Haggett1965}, there has been an understanding that regional organisation can be perceived through movement, and that the expansion of infrastructural networks is tightly linked to regional development. In this sense, the hierarchical perspective can be understood from the economic performance of regions, leaving traces in the road network's evolution.
The quest to characterise and quantify the regionalisation of the urban space in a hierarchical manner, dates back to the `30s. Initial ideas focused on subdividing the space, considering on the one hand morphological characteristics, and on the other population distribution. These were proposed in various contexts, from regular lattices \cite{Losch1938}, to  elaborated structures, such as the ones proposed by Christaller \cite{Christaller1933} in his Central-Place hierarchies.
Later on, efforts were directed into the hierarchical aspect of distance between settlements \cite{Losch1954,Brush_Bracey1955}, where a direct correlation between the size of a cluster and the distance to other clusters was identified \cite{Thomas1962}.
This was extended in the 60's, to a perspective on the functional characterisation of clusters according to size.
A hierarchy emerges with respect to the types of relationships that exist given the cluster size (whether the cluster is a village, a town or a city) \cite{Stafford1963,King1962,Berry_Garrison1958}.
These translate in contemporary urban studies to the study of scaling laws \cite{BettencourtScience2013, arcaute_Interface2015}, where nevertheless, it is assumed that all types emerge, and the hierarchy is indirectly linked with the aggregated value of the urban measurement, determined by the size of the cluster. It seems, that there is a need for a careful re-evaluation of the characterisation of urban indicators at various regimes of size.
Coming back to the hierarchical approach to urban functionalities, this has paved the way for different perspectives on urban modelling \cite{Burgess1927,Hoyt1939,Harris_Ullman1945,Berry_Pred1961,Wilson1974}.
Currently, network theory has been extremely successful and permeates the methodologies employed for many different urban models.
This approach nevertheless, dates back to the 60's, where graph theory was used to establish a hierarchical structure between regions, through the introduction of a measure of flow between places, from telephone calls to trade \cite{Nysten_Dacey1961}.

A recurrent aspect in the above mentioned approaches which decode urban hierarchies is the connectivity of the system. 
This can be explored through percolation theory \cite{Stauffer_Aharony_percolation1994,Bunde_Havlin1996}, which studies how a piece of information (or a disease, or a fire, etc.) spreads in space, reaching a critical point at which a giant cluster appears. 
In its most general form, the process is defined in an infinite lattice and for a random occupation probability.  
Relaxing these constraints, the analysis can be extended to finite systems, where the clusters are the outcome of some thresholding process. 
Some of these systems present a multiplicity of percolation transitions, revealing a hierarchical organisation. 
This was observed for the brain \cite{Gallos_Makse_brain2011}, where the percolation process is considered in terms of the connectivity between voxels given by the different stimuli thresholds.

A crude analogy can be drawn between the structure of the brain and that of an urban system. 
Both consist of highly integrated modules which connect to each other at different scales, giving rise to a functional system. 
For the urban system, the modules correspond to its cities, and its different regional divisions are a manifestation of its inherent hierarchical structure \cite{Haggett1965,Berry_Marble1968}.
We hence implement a similar methodology to \cite{Gallos_Makse_brain2011} on the street intersections of Britain, in order to unveil its hierarchical organisation.
Note that the network has been stripped away, and the percolation process is hence applied to the street intersections, which correspond to the occupied sites in space, connected to each other through proximity only. 
Using the intersection points as a proxy for urbanisation can be justified from archaeological times. In Anglo-Saxon Britain \cite{Brookes_Reynolds2011}, the assembly places were defined at the main points of convergence, where the relevant interactions took place. In contemporary times, it has been argued that the road intersections are the essential facilitators for the necessary human dynamics that lead to a productive urban system \cite{JaneJacobsCities}. In addition, there is also a technical bias that we purge ourselves from when removing the links of the network. This relates to the long-standing problems of digitisation of the dataset: faulty topology of the network, missing streets, disconnected networks given by the inaccuracy of streets almost meeting, etc. 
In any case, for the skeptical reader, we also provide a similar methodology developed directly on the entire street network, which has been carefully prepared and checked in order to avoid many of the above mentioned problems, and we show that the results are equally recovered.  

In the following sections we show that through a multiplicity of percolation transitions, the hierarchical structure of Britain emerges. 
These transitions indicate fractures of various sorts, from natural barriers, such as National Parks or lakes, to socio-economic polarisation such as the North-South divide.
The transition observed at the smallest scale defines the cores of the cities. 
It is well-known that the morphological properties of cities and regions are notably different. 
These have been extensively analysed for street networks \cite{Strano_Nicosia_Latora_Porta_Barthelemy2012,Porta_Crucitti_Latora2006_1,Porta_Crucitti_Latora2006_2,Barthelemy_Hauss2013,Masucci_PLosOne2013}, nevertheless, the statistical properties previously found cannot be used to define the boundaries of a city, since there is no clear transition between urban and rural networks. 
Here we show that through the analysis of the fractal dimension of the emergent clusters, a threshold can be identified at which cities are well-defined. 
This specific morphological property observed over the whole system, gives a maximum over all the thresholds. At this maximum, the obtained clusters are in very good correspondence with other proxies for cities, such as satellite images of the urbanised space, and previous definitions of cities proposed by the authors \cite{arcaute_Interface2015}.

\section{Methodology and dataset}

Percolation theory is classically approached in terms of the probability of a site being occupied in a lattice. It can also be thought of in terms of bond percolation, in which the sites are all occupied, and the probability corresponds to a bond to be open and to connect sites. 
In our analysis, the sites will correspond to the intersection points. 

In the following section we present two methodologies: 1) the percolation on the intersection points, and 2) the percolation on the street network. 
For both methods we use the most complete database for street networks in Britain: the Ordnance Survey (OS) MasterMap \cite{MasterMap2010}.
For computational purposes, we reduce the size of the dataset by introducing the following simplifications: 1) we remove the points that do not convey any morphological information, such as nodes of degree two, which for example correspond to streets changing name; 2) we replace roundabouts by a single intersection point, which is primarily relevant for the methodology on networks. The original dataset contains more than 23 million points and the final processed dataset contains only around 3 million intersection points. In detail, the network covering the whole of the UK has  $n=3,390,758$ nodes;  $m=3,973,186$ links; and the average node degree is $<k>=2.34$.

\subsection{Percolation on the street intersections}

For this method we take the dataset described above, and we remove all the street segments, leaving only the intersection points. 
We then apply a clustering algorithm that corresponds to a thresholding procedure parameterised by distance.
This is simply defined as the Euclidean distance between points, whether they are connected or not.
We observe different configurations of clusters appearing at different distances.
This procedure can be interpreted in terms of bond percolation as follows: the probability of a bond to be open between sites, corresponds to the distance between the intersection points. In this sense, one can think of a fully connected network in which the distance between nodes gives the probability for the link to exist after a normalisation procedure.

In practical terms, the algorithm is similar to the CCA (City Clustering Algorithm)  \cite{Rozenfeld_Batty_Makse08,Rozenfeld_Gabaix_Makse2011} based on population distribution in space, and the \emph{natural cities} definition given also in terms of road intersections \cite{Jiang_Jia2011}. In \cite{Rybski_percolation} this algorithm is also employed to understand the emergence of regions through percolation theory, and in \cite{Gallos_Makse_obesity2012}, in order to understand the spread of obesity in the United States.
It is important to note that most of these algorithms have been constructed in an effort to define cities in a consistent way, and considerable research is still undergoing in this direction  \cite{Rybski_PhysRevE87_2013,Frasco_etal_PhysRevX2014}.
These algorithms differ from models of urban growth based on correlated percolation \cite{Makse_Batty_Havlin_Stanley1998,Makse_Havlin_Stanley_Nature1995,Murcio_etal_Chaos2013}, and on correlations with urban sprawl \cite{Hernando2_Plastino2013}.

In detail, our algorithm is defined in terms of a distance parameter that determines clusters of intersection points in which every point has a neighbour at a distance equal or smaller than the given threshold. The algorithm can be implemented on the continuous space, or for large datasets requiring computationally demanding calculations, on a grid covering the space of points. Please refer to the appendix \ref{AlgPercolation} for more detail on the implementation of the algorithm.

\subsection{Percolation on the network}

In this case, we are considering the `real' network, where intersection points are connected if and only if there is a street connecting them. The clustering procedure is very similar to the procedure described above, but in this case the distance is given by the actual extent of the street. An open bond hence corresponds in this case to an existing street according to the different distance thresholds.
And once again, the links can be re-interpreted in terms of probabilities if the distances are normalised.
Details on how to implement this can be found in the appendix \ref{AlgPercolationNet}.

\section{Results}
\subsection{Urban hierarchies}

\begin{figure*}[ht]
\includegraphics[width=0.5\linewidth]{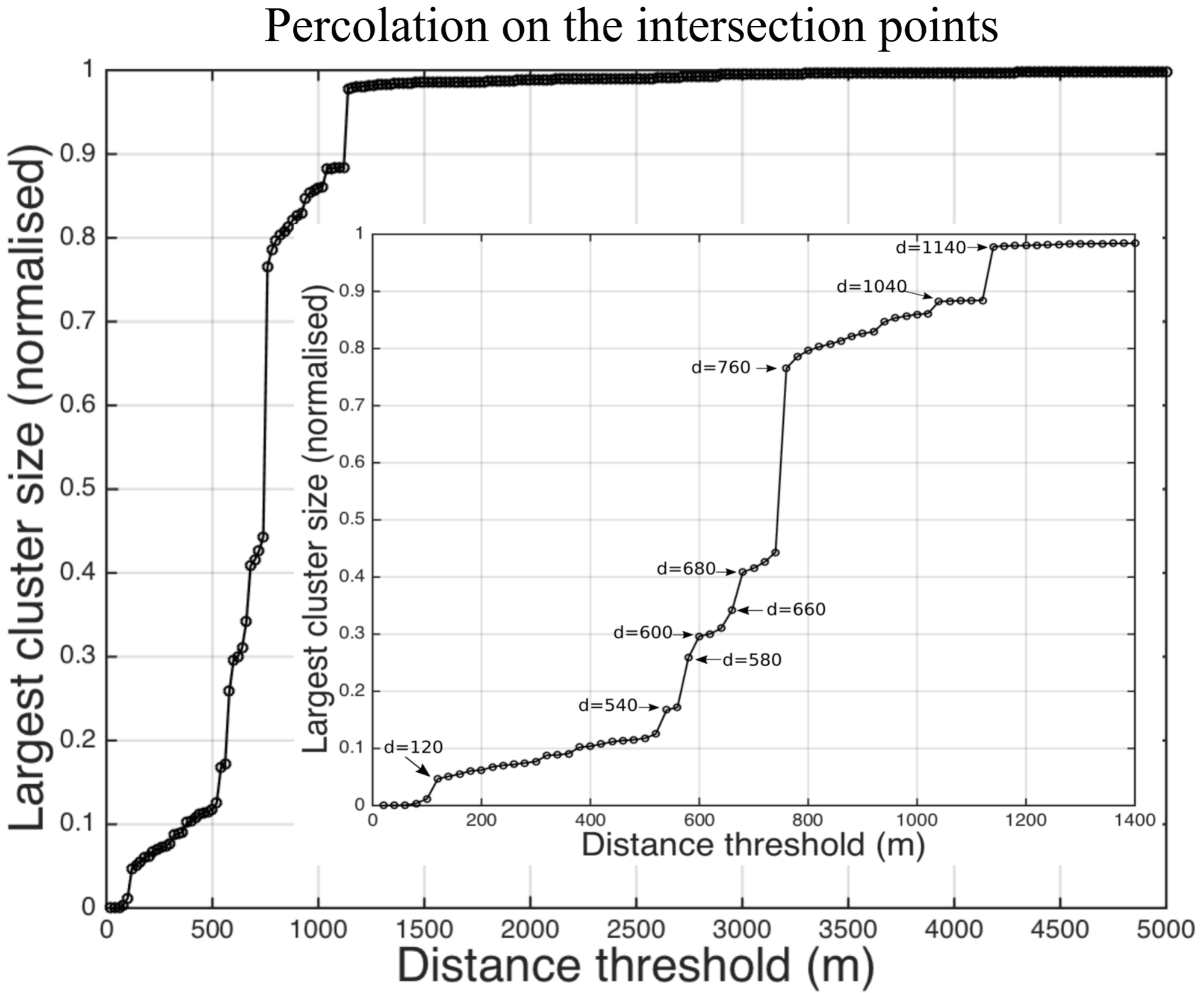}
\includegraphics[width=0.5\linewidth]{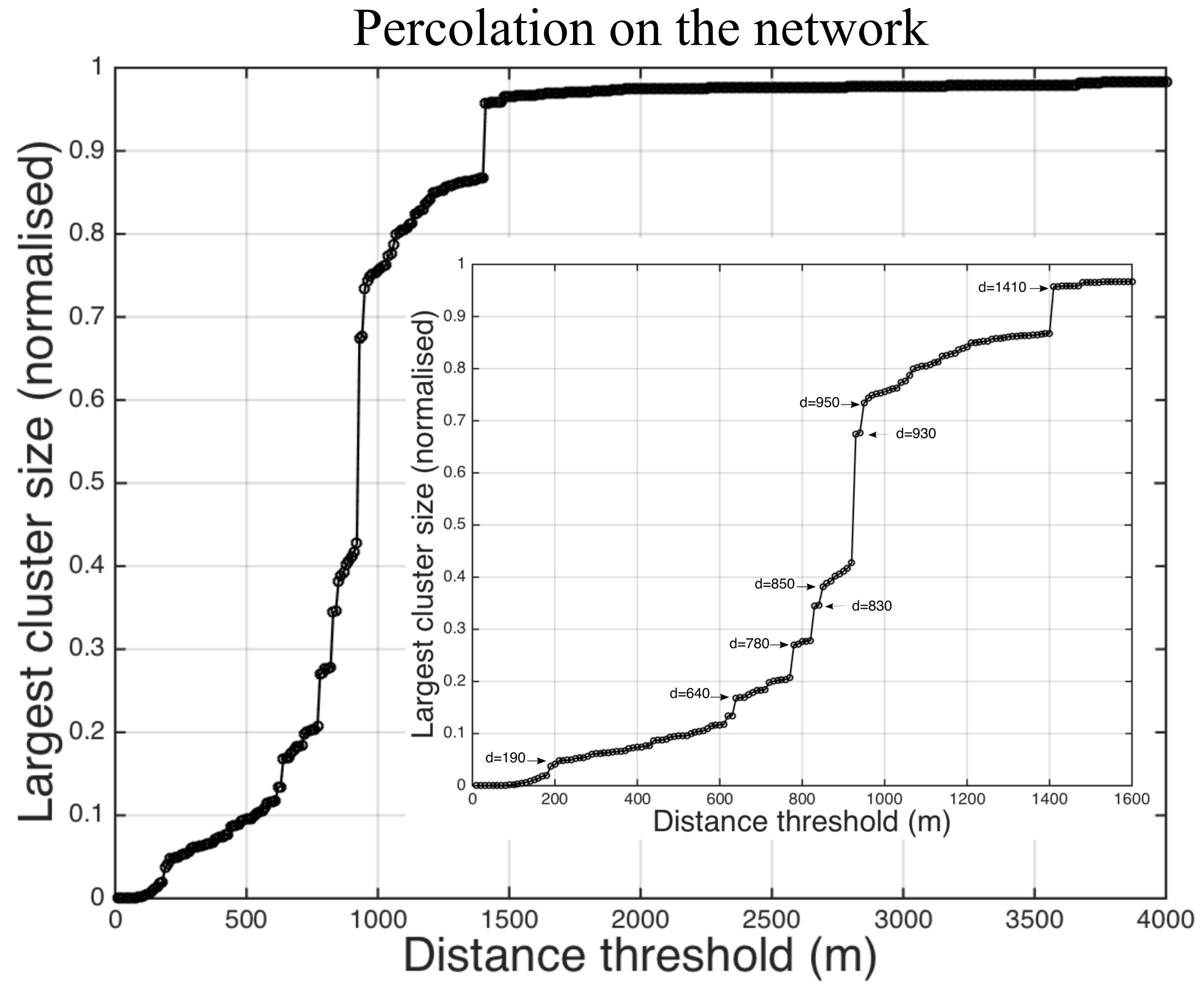}
\caption{Evolution of the largest cluster size for the percolation on both systems. The size has been normalised by the total number of intersection points present in the dataset.}\label{Smax_both}
\end{figure*}

We analyse the process in the traditional way, by looking at the size evolution of the largest cluster at the different distance thresholds \cite{Stauffer_Aharony_percolation1994}.
A multiplicity of percolation transitions defining the fractures at different scales is observed. The same divisions can be found in both systems, see Fig.~\ref{Smax_both}, although the critical distance threshold varies. 
As with any percolation process, the critical threshold that determines the transition pertains to the geometrical properties of the space. In the first system the percolation occurs on a regular grid, while in the second system, the process takes place on the network itself. 
We will present the rest of the results for the first scenario, hence all the critical distances correspond to those observed for the points of intersection, and are illustrated in Fig.~\ref{Smax_pts}. Details of the maps can be found in the appendix in Fig.~\ref{MapsGridPerc}.
\begin{figure*}[ht]
\centerline{\includegraphics[width=1\linewidth]{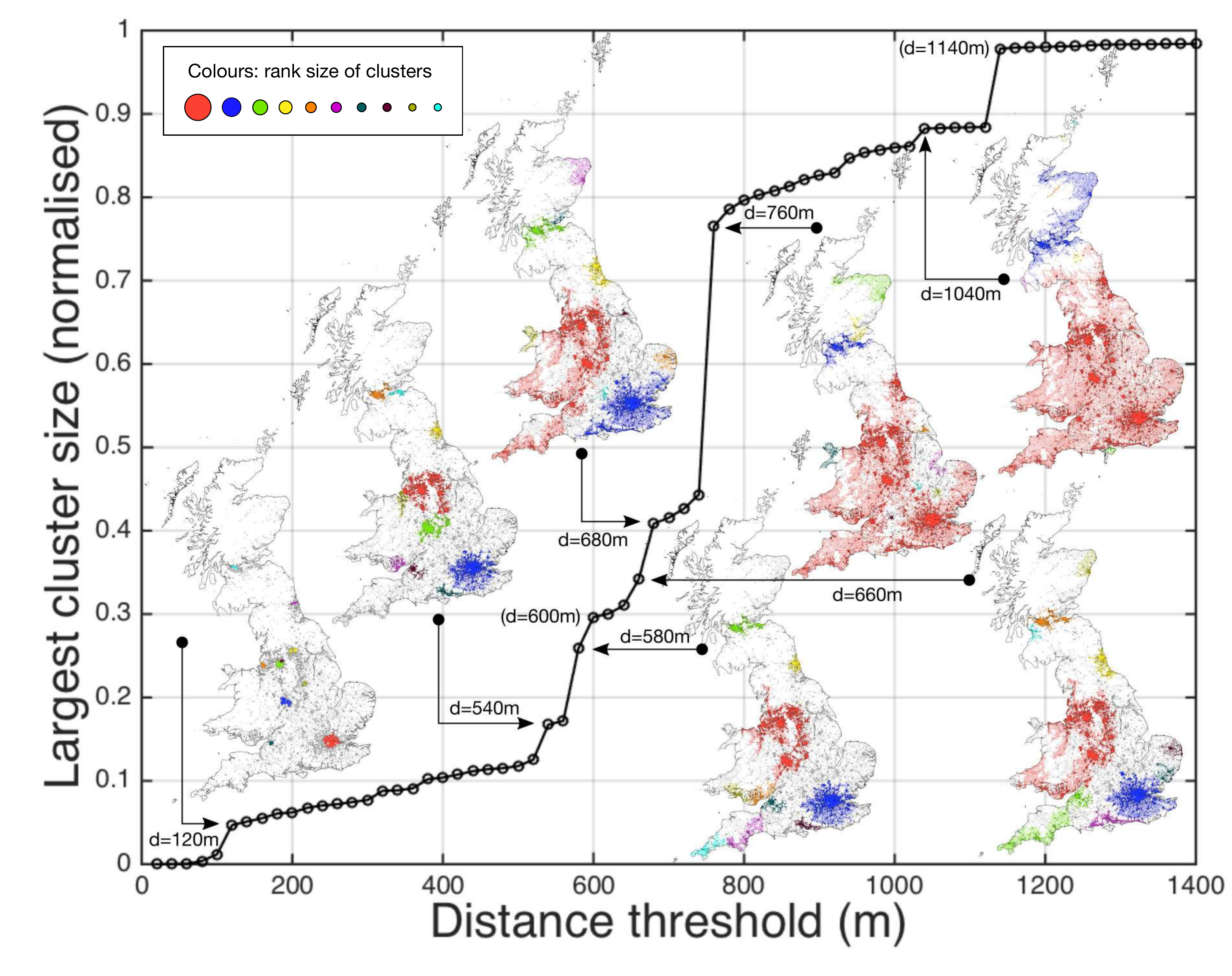}}
\caption{Evolution of the largest cluster size for the percolation on the intersection points.}\label{Smax_pts}
\end{figure*} 

The first transition detected at $d_c=120$m denotes the merging of the north and south parts of London separated by the river Thames. 
The giant cluster can be identified with the core of London. 
Looking across the system, other large clusters correspond to the cores of important cities, such as Birmingham, Manchester, Liverpool, Bristol, etc. This first transition is therefore representative for a system of cities. 
Nevertheless, although the cores are recovered, the extent of the cities is much smaller than expected. 
In the next section we will introduce a fractal analysis of the clusters at each distance threshold, and we will show that the system maximises the value of the fractal dimension at the highest point of correlation with the urbanised space, defined through a classification of satellite images. 

The giant cluster at the next transition, $d_c=540$m, encompasses the main post-industrial cities in the North: Liverpool, Manchester, Leeds and Sheffield. 
These are the main core cities conforming the region denominated as the ``Northern Powerhouse'' (devolution proposal)\footnote{This terminology adopted by the Chancellor of the Exchequer since June 2014 \cite{gov_Osborne_NP2014}, has been incorporated into the vocabulary of politics, featuring in the main news agencies in the UK \cite{BBC_NP2015,FT_NP2015}}, with some others, such as Newcastle missing in the cluster. 
The Northern Powerhouse proposal aims at devolving power and at boosting economic growth in cities in the North, reducing the gap of wealth between these and the cities in the South-East, mainly driven by London \cite{gov_NP2015}. 
Such a disparity in wealth distribution dates back to Roman times, and saw an intensification after de-industrialisation. Many claim that the reforms introduced by Margaret Thatcher's Conservative Governments during the 80's made the North-South divide even more significant.

The next transition at $d_c=580$m, sees an expansion of that region, introducing Birmingham into the largest cluster. 
The following smaller transitions see the annexing of Wales into the cluster at $d_c=660$m, and at $d_c=680$m, that of Cornwall.
At $d=740$m before the next transition takes place, a very clear division between the North-West and the South-East can be observed. It is important to note at this point that such a division is not the outcome of a geographical accident of sorts, say the presence of natural barriers such as parks, mountains or rivers. In addition, it is not an artifice of taking only the intersection points, since the split is also present when the percolation is performed directly on the network.
This split does not seem to be linked to the topography of the country, but to the division of wealth that we have been discussing throughout.
\begin{figure*}[ht]
\centering
	\includegraphics[width=1\linewidth]{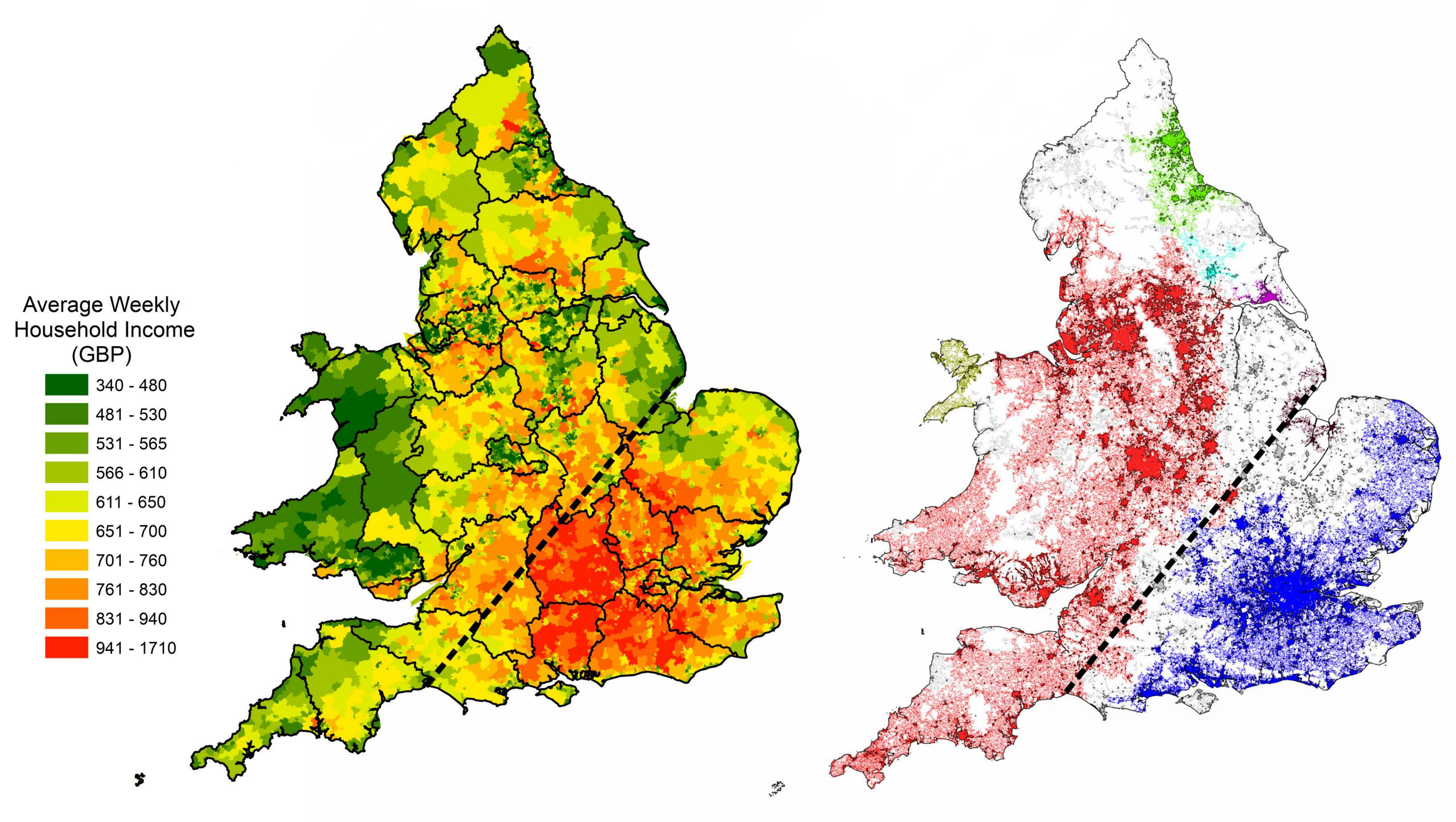}
\caption{Maps of England and Wales: right, at percolation distance threshold $d=740m$; left: thematic map of income with regional divisions given by NUTS2.}\label{NS_divide}
\end{figure*} 
We illustrate this in Fig.~\ref{NS_divide}. The map on the right shows the 10 largest clusters in colour at $d=740$m. 
The map on the left shows model-based income estimates for England and Wales at the level of Middle Layer Super Output Areas for 2007/08. These are based on census data, and hence Scotland is excluded from this map, since it has a different census to England and Wales. The black boundaries correspond to European administrative regional divisions called NUTS2 \cite{NUTS2}.
The dotted lines indicate a clear agreement between the clusters obtained from the percolation process, and the division of wealth in the country.

At $d_c=760$m, England and Wales, with the exception of Snowdonia (region of mountains and a National Park in Wales), merge into one region, marking the most striking transition.
This is followed by smaller transitions resulting from the merging of areas with natural barriers.
At $d=1120$m, Scotland can clearly be distinguished as a separate region from the rest of England and Wales.
Note that the split of Scotland from the rest of the country is not caused by barriers of topographic nature either.
This division is of a similar type to the one encountered earlier, it is the product of a historical cultural differentiation, that has been imprinted in the evolution of the infrastructure of the country.
The last important transition in the system is observed at $d_c=1140$m, and it corresponds to the merging of Scotland with England and Wales. 

It can be argued that these results are biased, given that the connectivity between points does not take into account whether these points can or cannot be connected through roads, since these were removed. 
To reassure the reader, we perform the percolation on the road network and we present the results in the appendix. 
The plots and maps, see Fig.~\ref{MapsNetPerc}, confirm the previous analysis, although at different critical distances.
These are larger, since they correspond to the length of the roads individuals need to take to travel from one intersection point to another in the urban system.
An important result of percolation theory, is that the value of the critical distances will vary from dataset to dataset, and from system to system. We hence do not expect to recover the same distances for the UK if a different dataset is used, nor the same distances for different countries.

\begin{figure*}[h]
\centering
	\includegraphics[width=1\linewidth]{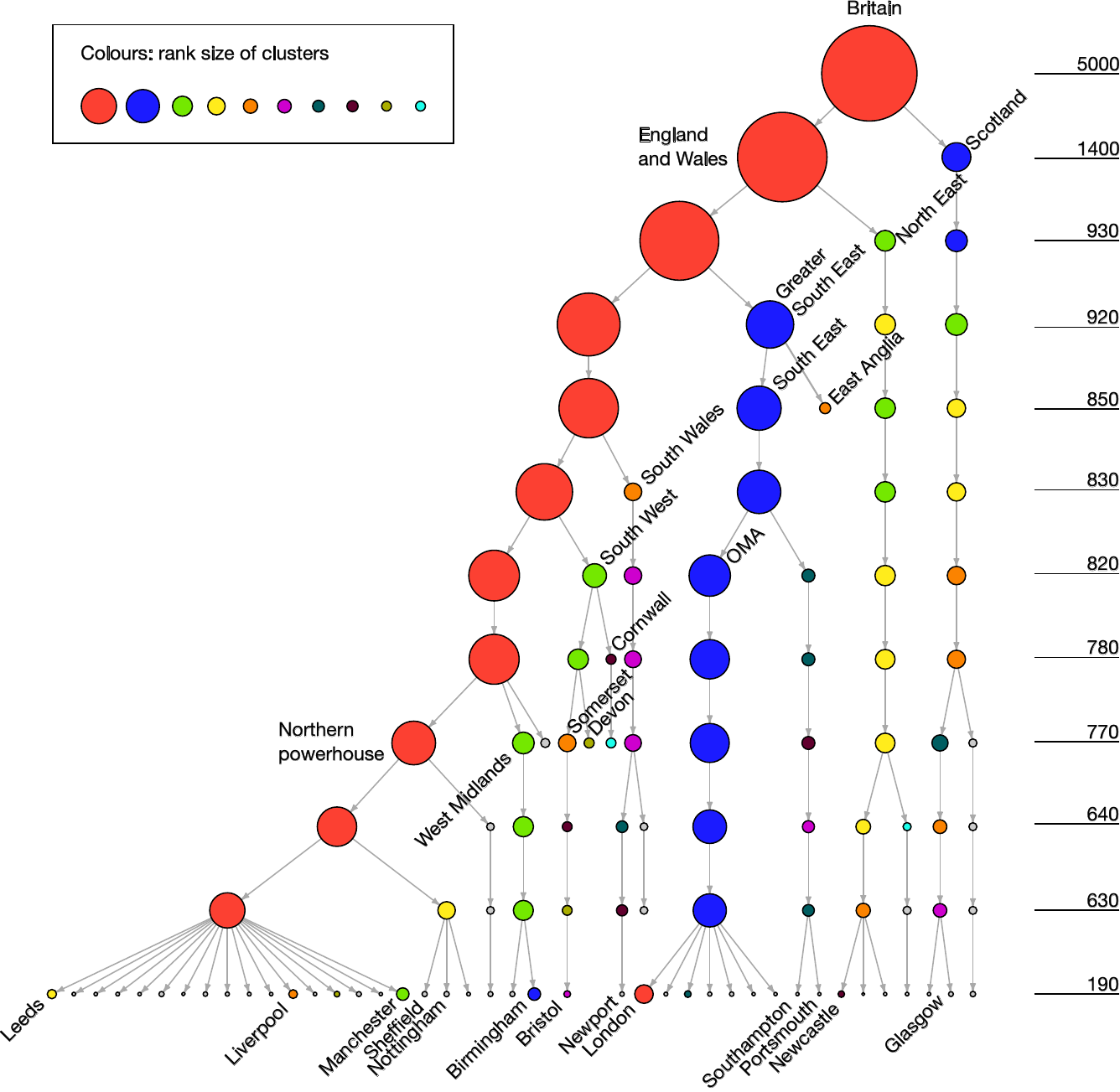}
\caption{Dendogram of the evolution of some of the largest clusters through the percolation process. The size is measured according to the number of intersection points.}\label{dendogram}
\end{figure*}

The results can be summarised in a crude way, by looking at the evolution of the relevant largest clusters through a dendogram, see Fig.~\ref{dendogram}. The hierarchical structure becomes very tangible, and the size of the nodes are scaled to represent the size of the real clusters.

\subsection{Fractal properties}

Urbanised spaces have specific morphological properties that cannot be found in non-urbanised areas, and many of these are manifested in the road network. 
We investigate in this section, whether the clusters defined at each distance threshold can characterise the urban system in a particular morphological manner.
We choose the fractal dimension as the property to analyse, given that this has been extensively researched for the morphological description of cities \cite{Batty_FractalCities,battylf1996preliminary,frankhauser1998fractal}. In \cite{Guerois_Pumain2008} for example, a fractal analysis was undertaken to compare built-up areas from the Corine dataset \cite{Corine_ref} at different built-up densities for 20 European cities.

This section contains two subsections for the analysis of each of the systems, since the fractal dimension cannot be computed in the same way for point patterns and for a network.

\subsubsection{Clusters of intersection points}

Until recently, the characterisation of the fractal structure of a system consisted of a single fractal dimension.
For cities, this was traditionally obtained through a box counting algorithm.
Nevertheless, this measure is extremely sensitive to the dataset and the implementation.
In addition, it is well recognised now, that cities are actually \emph{multifractals} \cite{chen2013multifractal,Murcio_Multifractal2015}. 
These are objects that present different fractal properties at different scales and regions \cite{stanley1988multifractal}, and hence cannot be fully characterised by a single fractal dimension, but need a spectrum of fractal dimensions \cite{appleby1996multifractal,ArizaVillaverde20131,Hu2012161}. 

\begin{figure}[H]
\centering
	\includegraphics[width=1\linewidth]{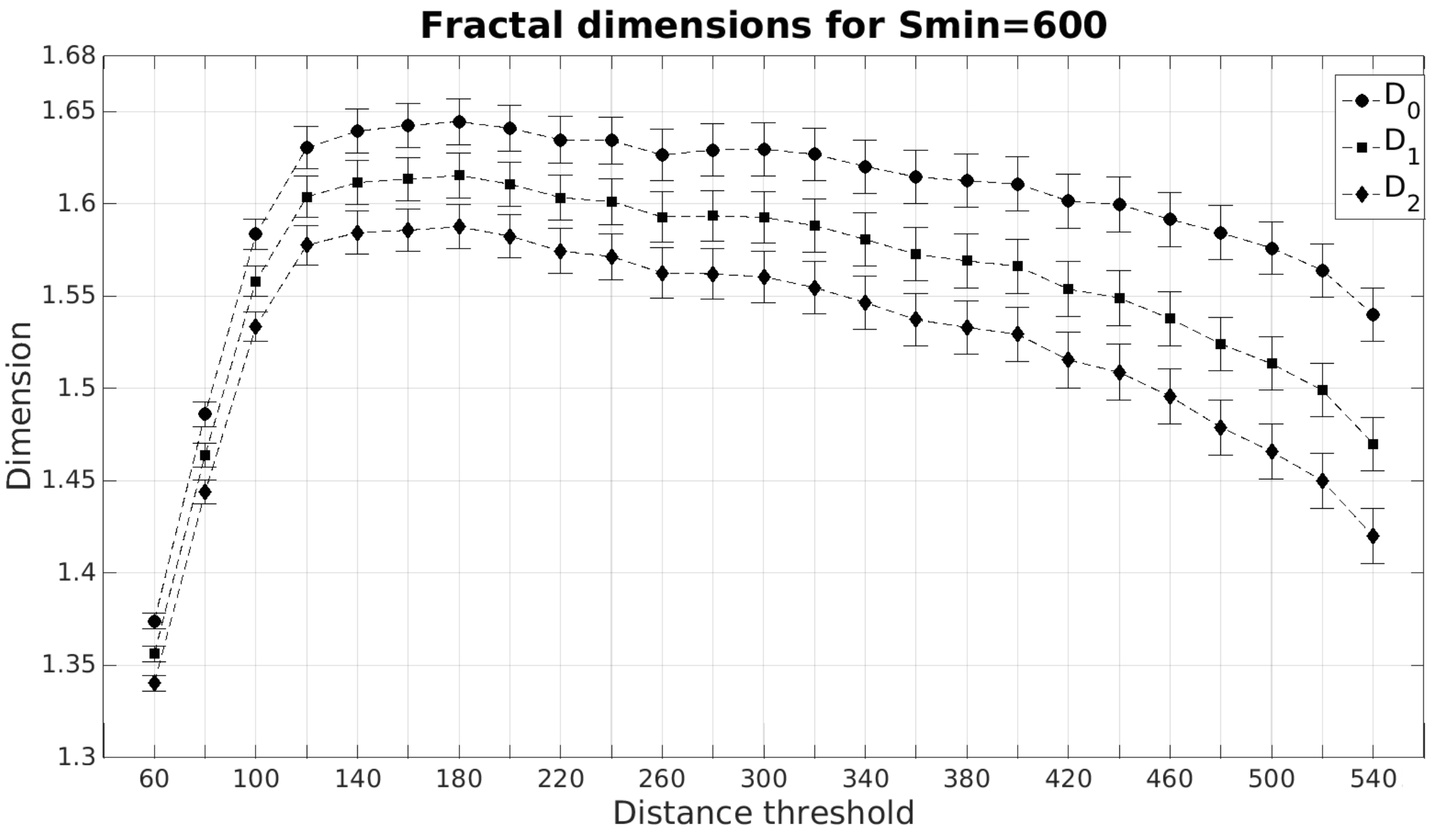}
\caption{Fractal spectrum of clusters with a minimum size of $S_{min}=600$ points obtained from the percolation on the intersection points.}\label{fractal_pts}
\end{figure}

\begin{figure*}[ht]
\centering
	\includegraphics[width=1\linewidth]{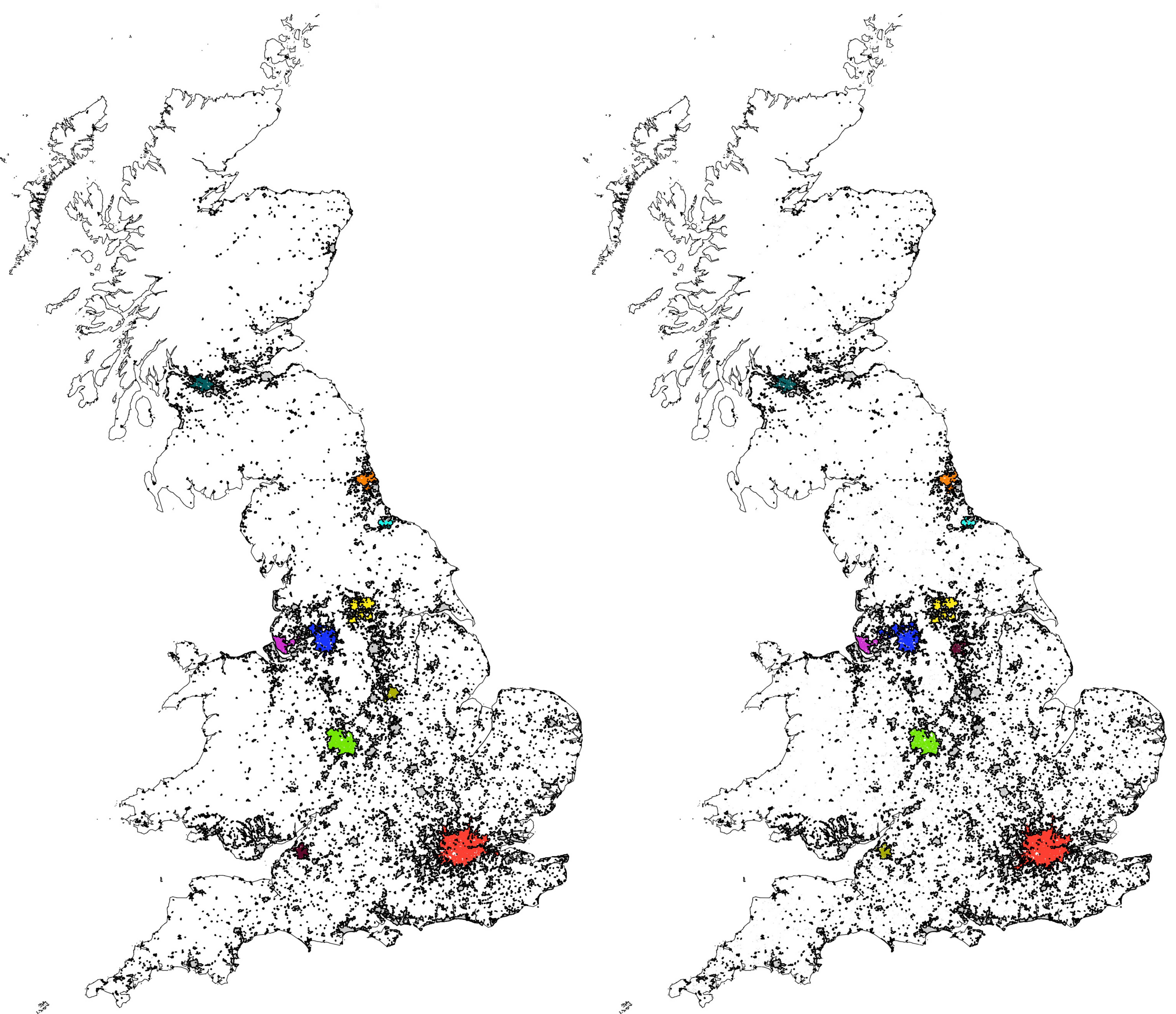}
\caption{All clusters appearing at the maximum fractal value. Left, for the intersection points at $d=180$m; right, for the network at $d=300$m, and black contours for the Corine dataset.}\label{mapsCities}
\end{figure*}

From the total spectrum, we select the three well-known fractal dimensions denoted by $D_0$, $D_1$ and $D_2$; where $D_0$ is the capacity dimension, and in practical terms it corresponds to the box-counting measure; $D_1$ is the information dimension and it can be interpreted as Shannon's entropy; and $D_2$ is the correlation dimension, which is considered to be the most accurate one. A quick review of these measures can be found in the appendix \ref{MultFractalReview}.
For the specific system at hand, we need to extract the characteristic fractal dimensions at each of the distance thresholds. 
Nevertheless, the clusters that emerge have all sorts of sizes, from individual points, to very large clusters.
Since we are interested in characterising the urban space, and ultimately in recovering the cities of the system, which are the landmark of urbanisation, we impose a minimum cluster size of 600 intersections for the computation of the fractal dimension.
In addition, we do not compute the fractal properties beyond a maximum distance threshold of $d=540$m, since the percolation method clearly returns regions beyond this transition, moving further and further away from a configuration of cities. 
The methodology to compute these three dimensions follows the same algorithms described in \cite{Murcio_Multifractal2015}, where further details can be found. 
The results can be seen in Fig.~\ref{fractal_pts}.
The evolution of all the three fractal properties of the system, indicates a maximum at $d=180$m.
Carefully inspecting the urban system at this maximum of its morphological characterisation, we observe that the clusters at this distance threshold are in excellent correspondence with the classification of urbanised space given by the Corine dataset \cite{Corine_ref}, which is a classification of the Landsat satellite images. 
The high level of agreement can be seen in the Fig.~\ref{mapsCities}, where the colour of the clusters are chosen according to size, and the black contours correspond to the classified urbanised areas.

\subsubsection{Clusters of networks}

Let us now compute the fractal dimension of the clusters of networks that emerge from the percolation on the network.
In this case, the fractal dimension $\alpha$ of the system is computed in terms of the scaling relationship between the mass of the clusters and the diameter of the network. The mass is given by the number of intersection points $N$ and the diameter is denoted by $r_{max}$, leading to
\begin{equation}
N\sim r_{max}^{\alpha}\label{N_rmax}
\end{equation}

This corresponds to the same methodology implemented in \cite{Gallos_Makse_brain2011}. Note that for this system we need to take a slightly larger maximum distance threshold $d=800m$ to ensure that we are well within the cities definition. 
In addition, in order to include as many small settlements as possible in the analysis, we use for networks a minimum cluster size of $S_{min}=50$ nodes (a node is an intersection point), instead of 600.

\begin{figure}[H]
\centering
	\includegraphics[width=1\linewidth]{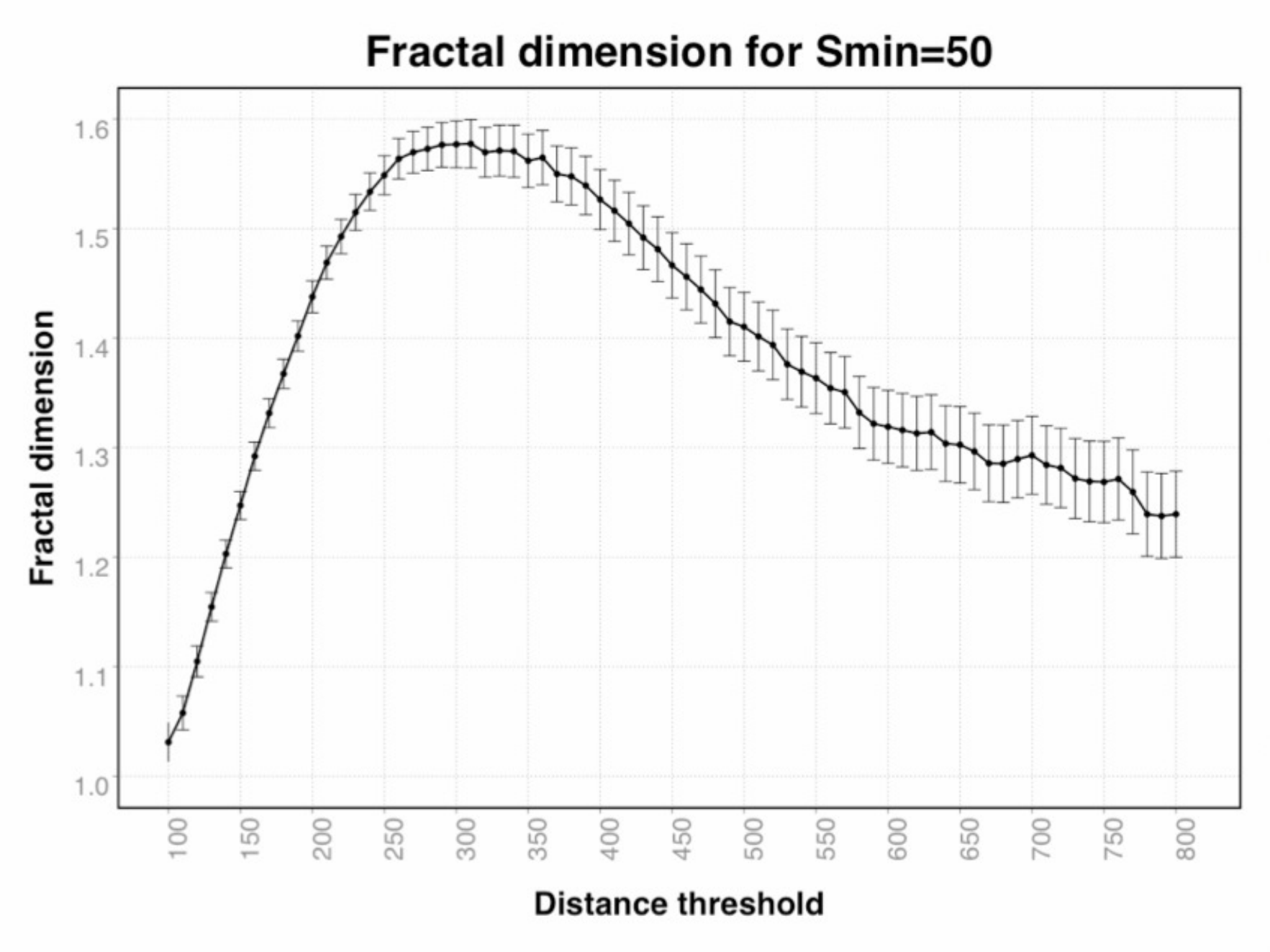}
\caption{Fractal dimension of the whole urban system. It is computed on the networks obtained at different distance thresholds, using the scaling relationship between mass and diameter given by Eq.(\ref{N_rmax}).}\label{fractal_NTs}
\end{figure}

For this case the results show a maximum at around $d=300$m, see Fig.~\ref{fractal_NTs}.
And once again we see that the urban system defined at this maximum is in excellent correspondence with the definition of cities. Fig.~\ref{mapsCities} shows the clusters at $d=300$m, with the contours for the classified urbanised areas in black.

\begin{figure}[H]
\centering
	\includegraphics[width=1\linewidth]{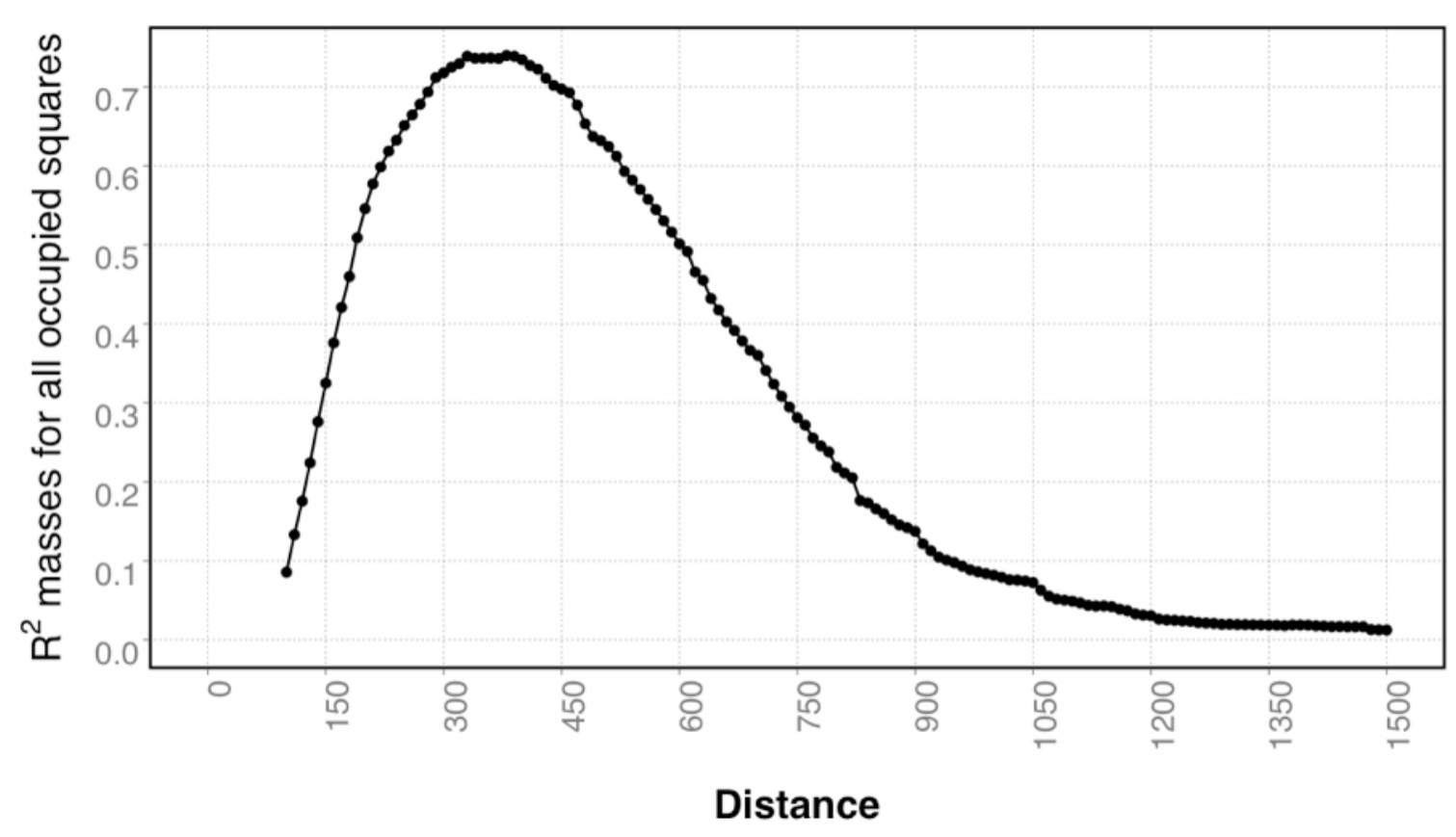}
\caption{Correlation of the clusters from the network percolation with the boundaries of the Corine dataset.}\label{R2_Corine_NTS}
\end{figure}

We quantify the level of agreement between the clusters obtained through the percolation method and the urbanised areas in the Corine dataset through a correlation measure. Fig.~\ref{R2_Corine_NTS} indicates that the maximum of this correlation is also in the vicinity of $d=300$m. Further details of how this measure is computed can be found in the appendix \ref{CorrCorine}.

It is important to reiterate that the distance is not universal nor uniquely characterised. It is not universal, because it depends on the nature of the dataset. Hence a distance of $d=300$m might suit this specific dataset for Britain, but might not suit another dataset, nor another European country. It is not uniquely defined, because the maximum corresponds to some sort of plateau. Hence any definition in the vicinity of $d=300$m would be as accurate or as inaccurate as the one for $d=300$m.

\section{Conclusions}

Throughout this work we have shown that percolation theory can be applied to the street intersections and network, in order to uncover its intrinsic hierarchical structure. 
We argued that such an organisation does not only relate to ideological or geographical divisions, but it represents a socio-economic polarisation of the system. 
It has been extensively discussed that the regional development is tightly linked to the development of its infrastructure, hence it should not be surprising to find these economic patterns reflected in the density of the road network.

Regions are formed by settlements which share a stronger connectivity among themselves than among settlements of other regions. It is therefore not surprising to recover historical trajectories, and existing alliances.
In this sense, looking at the highly integrated region of the cities in the North at $d=540$m, it appears that the assimilation of Newcastle within the Northern Powerhouse proposal needs to be done with care, so that it is not left behind, given its weaker connectivity to the rest of the cities in the region.
In this sense, many regional policies need to consider the strong ties that lie within the urban system.

From the perspective of the methodology, this formalism has the advantage that it can be implemented for incomplete datasets.
We would have to test the robustness of the method with respect to the incompleteness of the dataset, but at this stage it is clear that the road intersections serve as a good proxy for urbanisation \cite{Garrison_highway1959}, and that the percolation process on the point patterns recovers the hierarchical organisation of the system. 
It is to be expected that the level of detail provided by the dataset would affect the level of detail for some of the transitions in the system, nevertheless, a hierarchical sketch can still be recovered.
In addition, we have extrapolated the method to other spatial distributions where data are sparse, such as census data from the eleventh century, i.e. data from the Domesday Book.

Finally, this work has also provided a framework to define boundaries of cities in a global way, using a dataset that is open and not constrained to geographical units, such as the census data. 
In previous work we developed a procedure to define cities using population density from the census \cite{arcaute_Interface2015}.
Although successful, this procedure relies very heavily on data only available every 10 years, and on the level of granularity of the geographical unit. Nevertheless, it is still a useful method, since through the use of commuting data functional areas beyond urban cores can be defined, such as metropolitan areas.
Note that a further refinement of the percolation approach can be found in \cite{Masucci_CondThresh}, where each city is adjusted to its condensation threshold. 

Future research is needed to understand the mechanisms that drive the system to a maximum fractal dimension at the point where the cities reach their urban extent.

\section*{Acknowledgment}

EA, CM, RM, PM and MB acknowledge the support of ERC Grant 249393-ERC-2009-AdG.
EH acknowledges support from J. M. Epstein's NIH Director's Pioneer Award, number DP1OD003874 from the National Institute of Health.
We would also like to acknowledge useful discussions with Hern\'an Makse. 


\begin{thebibliography}{10}

\bibitem{MolineroVoting_arXiv2015}
Molinero C, Arcaute E, Smith D, Batty M. 2015.
 The Fractured Nature of {B}ritish Politics.
\textit{arXiv:150500217 [physics.soc-ph]}.

\bibitem{Haggett1965}
Haggett P. 1965.
\textit{Locational Analysis in human geography}.
London: Edward Arnold.

\bibitem{Losch1938}
L{\"o}sch A. 1938.
The nature of economic regions.
\textit{S Econ J.} 5:71--8.

\bibitem{Christaller1933}
Christaller W. 1966.
\textit{Central Places in Southern Germany}.
 Englewood Cliffs, NJ: Prentice-Hall. Original work published in
  1933 as \emph{Die Zentrale Orte in Suddeutschalnd}, Jena, Germany: Gustav
  Fisher.

\bibitem{Losch1954}
L{\"o}sch A. 1954.
\textit{The economics of location}.
 New Haven, CT: Yale University Press

\bibitem{Brush_Bracey1955}
Brush JE, Bracey HE. 1955.
 Rural service centres in southwestern {W}insconsin.
\textit{Geographical Review} 43:380--402.

\bibitem{Thomas1962}
Thomas EN. 1962.
 The stability of distance-population size relationships for {I}owa
  towns from 1900 to 1950.
\textit{ Lund Studies in Geography, Series B, Human Geography} 24:13--30.

\bibitem{Stafford1963}
{Stafford HA Jr }. 1963.
 {The functional bases of small towns}.
\textit{ Economic geography} 39(2):165--175.

\bibitem{King1962}
King LJ. 1962.
 A Quantitative Expression of the Pattern of Urban Settlements in
  Selected Areas of the United States.
\textit{ Tijdschrift voor Economische en Sociale Geografie} 53:1--7.

\bibitem{Berry_Garrison1958}
Berry BJL, Garrison WL. 1958.
 Functional bases of the central place hierarchy.
\textit{ Economic Geography} 34:145--154.

\bibitem{BettencourtScience2013}
Bettencourt LMA. 2013
 The origins of scaling in cities.
\textit{ Science} 340:1438--1441.

\bibitem{arcaute_Interface2015}
Arcaute E, Hatna E, Ferguson P, Youn H, Johansson A, Batty M. 2014
 Constructing cities, deconstructing scaling laws.
\textit{ J R Soc Interface} 12(102).

\bibitem{Burgess1927}
Burgess EW. 1927.
 The determination of gradients in the growth of the city.
\textit{ American Sociological Society} 21:178--84.

\bibitem{Hoyt1939}
Hoyt H. 1939.
\textit{ The structure and growth of residential neighbourhoods in {A}merican cities}.
 Washington, D.C.: Federal Housing Administration.

\bibitem{Harris_Ullman1945}
Harris CD, Ullmann EL. 1945.
 The nature of cities.
\textit{ Annals of the American Academy of Political Science} 242:7--17.

\bibitem{Berry_Pred1961}
Berry BJL, Pred A. 1961.
\textit{ Central place studies: a bibliography of theory and applications}.
 Regional Science Research Institute, Bibliographic Series I.

\bibitem{Wilson1974}
Wilson AG. 1974.
\textit{ Urban and regional models in geography and planning}.
 London: John Wiley and Sons.

\bibitem{Nysten_Dacey1961}
Nystuen JD, Dacey MF. 1961.
 A graph theory interpretation of nodal regions.
\textit{ Papers in Regional Science} 7(1):29--42.

\bibitem{Stauffer_Aharony_percolation1994}
Stauffer D, Aharony A. 1994.
\textit{ Introduction to Percolation Theory}.
 London: Taylor and Francis.

\bibitem{Bunde_Havlin1996}
Bunde A, (Editors) SH. 1996.
\textit{ Fractals and Disordered Systems}.
 Berlin: Springer.

\bibitem{Gallos_Makse_brain2011}
Gallos LK, Makse HA, Sigman M. 2012.
 A small world of weak ties provides optimal global integration of
  self-similar modules in functional brain networks.
\textit{ PNAS} 109(8):2825--2830.

\bibitem{Berry_Marble1968}
Berry BJL, (Editors) DFM. 1968.
\textit{ Spatial Analysis: a reader in statistical geography}.
 New Jersey: Prentice-Hall.

\bibitem{Brookes_Reynolds2011}
Brookes S, Reynolds A. 2011 The origins of political order and the Anglo-Saxon State, \textit{Archaeol. Int.} 13, 84–93. (doi:10.5334/ai.1312)


\bibitem{JaneJacobsCities}
Jacobs J. 1961.
\textit{ The Death and Life of Great American Cities}.
 New York: Random House.

\bibitem{Strano_Nicosia_Latora_Porta_Barthelemy2012}
Strano E, Nicosia V, Latora V, Porta S, Barthelemy M. 2012
 Elementary processes governing the evolution of road networks.
\textit{ Sci Rep.} 2.

\bibitem{Porta_Crucitti_Latora2006_1}
Porta S, Cruciti P, Latora V. 2006.
 The network analysis of urban streets: a primal approach.
\textit{ Env Plan B.} 33(5):705--725.

\bibitem{Porta_Crucitti_Latora2006_2}
Porta S, Cruciti P, Latora V. 2006.
 The network analysis of urban streets: A dual approach.
\textit{ Phys A.} 369:853--866.

\bibitem{Barthelemy_Hauss2013}
Barthelemy M, Bordin P, Berestycki H, Gribaudi M. 2013.
 Self-organization versus top-down planning in the evolution of a
  city.
\textit{ Sci Rep.} 3(2153).

\bibitem{Masucci_PLosOne2013}
Masucci P, Stanilov K, Batty M. 2013.
 Limited Urban Growth: London's Street Network Dynamics since the 18th
  Century.
\textit{ PLoS One.} 8(8):e69469.

\bibitem{MasterMap2010}
OS MasterMap Integrated Transport Network Layer 
 [GML geospatial  data], Coverage: Great Britain, Updated Jan 2010, Ordnance Survey, GB. Using:  EDINA Digimap Ordnance Survey Service.

\bibitem{Rozenfeld_Batty_Makse08}
Rozenfeld HD, Rybski D, Andrade JS Jr, Batty M, Stanley HE, Makse HA. 2008.
 Laws of population growth.
\textit{ PNAS} 105(48):18702--18707.

\bibitem{Rozenfeld_Gabaix_Makse2011}
Rozenfeld HD, Rybski D, Gabaix X, Makse HA. 2011.
 The area and population of cities: new insights from a different
  perspective on cities.
\textit{ Am Econ Rev.} 101:2205--2225.

\bibitem{Jiang_Jia2011}
Jiang B, Jia T. 2011.
 Zipf's law for all the natural cities in the United States: a
  geospatial perspective.
\textit{ Int J of Geo Inf Sci.} 25(8):1269--1281.

\bibitem{Rybski_percolation}
Fluschnik T, Kriewald S, Garc{\'i}a Cant{\'u}~Ros A, Zhou B, Reusser D, Kropp J, et~al. 2014.
 The size distribution, scaling properties and spatial organization of  urban clusters: a global and regional perspective.
\textit{ arXiv:14040353 [physics.soc-ph]}.

\bibitem{Gallos_Makse_obesity2012}
Gallos LK, Bartfeld P, Havlin S, Sigman M, Makse HA. 2012.
 Collective behavior in the spatial spreading of obesity.
\textit{ Sci Rep.} 2(454):1--9.

\bibitem{Rybski_PhysRevE87_2013}
Rybski D, Garc{\'i}a Cant{\'u}~Ros A, Kropp JP. 2013.
 Distance-weighted city growth.
\textit{ Phys Rev E.} 87:042114.

\bibitem{Frasco_etal_PhysRevX2014}
Frasco GF, Sun J, Rozenfeld HD, ben Avraham D. 2014.
 Spatially Distributed Social Complex Networks.
\textit{ Phys Rev X.} 4:011008.

\bibitem{Makse_Batty_Havlin_Stanley1998}
Makse HA, Andrade JS, Batty M, Havlin S, Stanley HE. 1998.
 Modelling urban growth patterns with correlated percolation.
\textit{ Phys Rev E.} 58(6):7054--7062.

\bibitem{Makse_Havlin_Stanley_Nature1995}
Makse HA, Havlin S, Stanley HE. 1995.
 Modelling urban growth patterns.
\textit{ Nature} 377:608--612.

\bibitem{Murcio_etal_Chaos2013}
Murcio R, Sosa-Herrera A, Rodriguez-Romo S. 2013.
 Second order metropolitan urban phase transitions.
\textit{ Chaos, Solitons and Fractals} 48:22--31.

\bibitem{Hernando2_Plastino2013}
Hernando A, Hernando R, Plastino A. 2013.
 Space{\textendash}time correlations in urban sprawl.
\textit{ J R Soc Interface} 11(91).

\bibitem{gov_Osborne_NP2014}
Chancellor: 'We need a Northern powerhouse' (speech transcript).
\emph{Gov.uk}. Government of the United Kingdom. 23 June 2014.
  Retrieved 25 January 2016. See
\url{https://www.gov.uk/government/speeches/chancellor-we-need-a-northern-powerhouse}.

\bibitem{BBC_NP2015}
What is the Northern Powerhouse?.
\emph{BBC News}. 14 May 2015. Retrieved 25 January 2016. See \url{http://www.bbc.co.uk/news/magazine-32720462}.

\bibitem{FT_NP2015}
Concerns for Northern Powerhouse after head of tech body quits.
\emph{The Financial Times}. 25 January 2016. Retrieved 25 January
  2016. See \url{http://on.ft.com/1ZNYv49}.

\bibitem{gov_NP2015}
The Northern Powerhouse: One Agenda, One Economy, One North;.
\emph{Gov.uk}. Department for Transport. March 2015. Retrieved 25
  January 2016.

\bibitem{NUTS2}
NUTS2. Office for National Statistics. Nomenclature of Territorial Units for
  Statistics ({NUTS}) / {L}ocal {A}dministrative {U}nits ({LAU}). See \url{http://www.ons.gov.uk/ons/guide-method/geography/beginner-s-guide/eurostat/index.html}.

\bibitem{Batty_FractalCities}
Batty M, Longley P. 1994.
\textit{ Fractal Cities: A Geometry of Form and Function}.
 Academic Press, San Diego, CA and London.

\bibitem{battylf1996preliminary}
Batty M, Xie Y. 1996.
 Preliminary evidence for a theory of the fractal city.
\textit{ Environment and Planning A.} 28:1745--1762.

\bibitem{frankhauser1998fractal}
Frankhauser P. 1998.
\textit{ The fractal approach. A new tool for the spatial analysis of urban agglomerations}.
 Population: An English Selection, p. 205--240.

\bibitem{Guerois_Pumain2008}
Gu{\'e}rois M, Pumain D. 2008.
Built-Up Encroachment and the Urban Field: A Comparison of Forty
  European Cities.
\textit{ Environment and Planning A.} 40(9):2186--2203.
Available from:
  \url{http://epn.sagepub.com/content/40/9/2186.abstract}.

\bibitem{Corine_ref}
EEA. 2002.
Corine land cover update 2000, technical guidelines.
Copenhagen: European Environment Agency. ISBN: 92-9167-511-3.

\bibitem{chen2013multifractal}
Chen Y, Wang J. 2013.
 Multifractal characterization of urban form and growth: the case of  Beijing.
\textit{ Env Plan B.} 40(5):884--904.

\bibitem{Murcio_Multifractal2015}
Murcio R, Masucci AP, Arcaute E, Batty M. 2015.
 Multifractal to monofractal evolution of the London's street network.
\textit{Phys Rev E.} 92(6):062130.

\bibitem{stanley1988multifractal}
Stanley HE, Meakin P, et~al. 1988.
 Multifractal phenomena in physics and chemistry.
\textit{ Nature} 335(6189):405--409.

\bibitem{appleby1996multifractal}
Appleby S.  1996.
 Multifractal characterization of the distribution pattern of the  human population.
\textit{ Geographical Analysis} 28(2):147--160.

\bibitem{ArizaVillaverde20131}
Ariza-Villaverde AB, Jim{\'e}nez-Hornero FJ, Rav{\'e} EGD. 2013.
 Multifractal analysis of axial maps applied to the study of urban morphology.
\textit{ Computers, Environment and Urban Systems} 38(0):1 -- 10.

\bibitem{Hu2012161}
Hu S, Cheng Q, Wang L, Xie S. 2012.
 Multifractal characterization of urban residential land price in  space and time.
\textit{ Applied Geography} 34(0):161 -- 170.

\bibitem{Garrison_highway1959}
Garrison ML, Berry BJ, Marble DF, Nystuen JD, Morrill RL. 1959.
Studies of highway development and geographic change.
Seattle, WA: University of Washington Press.

\bibitem{Masucci_CondThresh}
Masucci P, Arcaute E, Hatna E, Stanilov K, Batty M. 2015.
 On the problem of boundaries and scaling for urban street networks.
\textit{ J R Soc Interface} 12(20150763).

\bibitem{Clauset_etalPL09}
Clauset A, Shalizi CR, Newman MEJ. 2009.
Power-Law Distributions in Empirical Data.
\textit{ SIAM Review.} 51(4):661--703.

\end{thebibliography}


\clearpage

\renewcommand{\thefigure}{S\arabic{figure}}
\renewcommand{\thetable}{S\arabic{table}}
\setcounter{figure}{0}


\onecolumn 

\section*{Appendix}

\renewcommand{\thesubsection}{\Alph{subsection}}

\subsection{Algorithm for the percolation on the intersection points}\label{AlgPercolation}

The algorithm is based on the geographical location of intersections. We consider a pair of intersections as connected if they are no more than $d$ meters apart. In order to reduce the computational complexity of the procedure, the actual analysis is performed using a grid of squared cells ($10\times10$ meters each). A cell has one of two values:  1 if at least one intersection is within its area or \textit{null} if it contains no intersections.  As the percolation analysis is based on distance, we calculate a distance grid where each cell is assigned the distance to the closest cell that contains an intersection. We use this grid in the percolation procedure.

The percolation procedure for a distance $d$ consists of the following steps:
\begin{enumerate}
\item
Each cell of the distance grid that has a distance value of $d$ meters or below is marked as 1, otherwise, it is marked as \textit{null}.
\item
A unique identifier is assigned to each continuous set of marked cells.  A cell is considered adjacent to its four nearest neighbours (i.e., its von Neumann neighbour).
\item
Each intersection is assigned the unique identifier of its containing cell. 
\end{enumerate}

The method is implemented in ESRI ArcMap 10.1 using the following tools:
\begin{itemize}
\item
The intersection grid is created using the \textit{Points to Raster} tool.
\item
The distance grid is created using the \textit{Euclidian distance} tool.
\item
The marked cells grid is created using the  \textit{Raster Calculator} tool.
\item
The unique identifiers grid is created using the \textit{Region Group} tool.
\item
The unique identifiers are copied to the intersection points using the \textit{Extract Values to Points} tool.
\end{itemize}

\subsection{Algorithm for the network based percolation}\label{AlgPercolationNet}

Given a graph of the road network, where nodes represent intersections and the weight for each edge is the length of the street that connects them and a certain metric threshold (e.g. 5000m) we produce a network percolation via the following steps:
\begin{enumerate}
\item We select the link of the graph with the smallest weight (distance), generating a new cluster and inserting both its nodes into the cluster.
\item We keep a first-in first-out queue of {\it nodes to expand}, from which we extract a node to continue the process. We add both nodes of the link selected in step 1 to this queue.
Nodes are only added to this queue if they are not already included.
\item We extract a node from the queue of {\it nodes to explore} and if a link departing from that node (not yet included in the cluster) is smaller than the threshold, include the link in the cluster and the end node of the link in the queue of nodes to explore.
\item We repeat step 3 until no further node can be expanded (the queue is empty) and if there are links left in the graph that do not belong to any cluster, generate a new cluster by choosing the smallest available link and repeat from step 1. 
\end{enumerate}

\subsection{Details of the clusters at the transitions}

At each of the critical distances defining a transition, a set of clusters appears. 
There can be thousands of these, hence we opt to only visualise the 10 largest ones, setting colours representing a rank size: red is the biggest, blue is the second biggest, green the third etc. The rank and colour is illustrated in Fig.~\ref{dendogram}.

\begin{figure}[ht]
\centering
	\includegraphics[height=0.95\textheight]{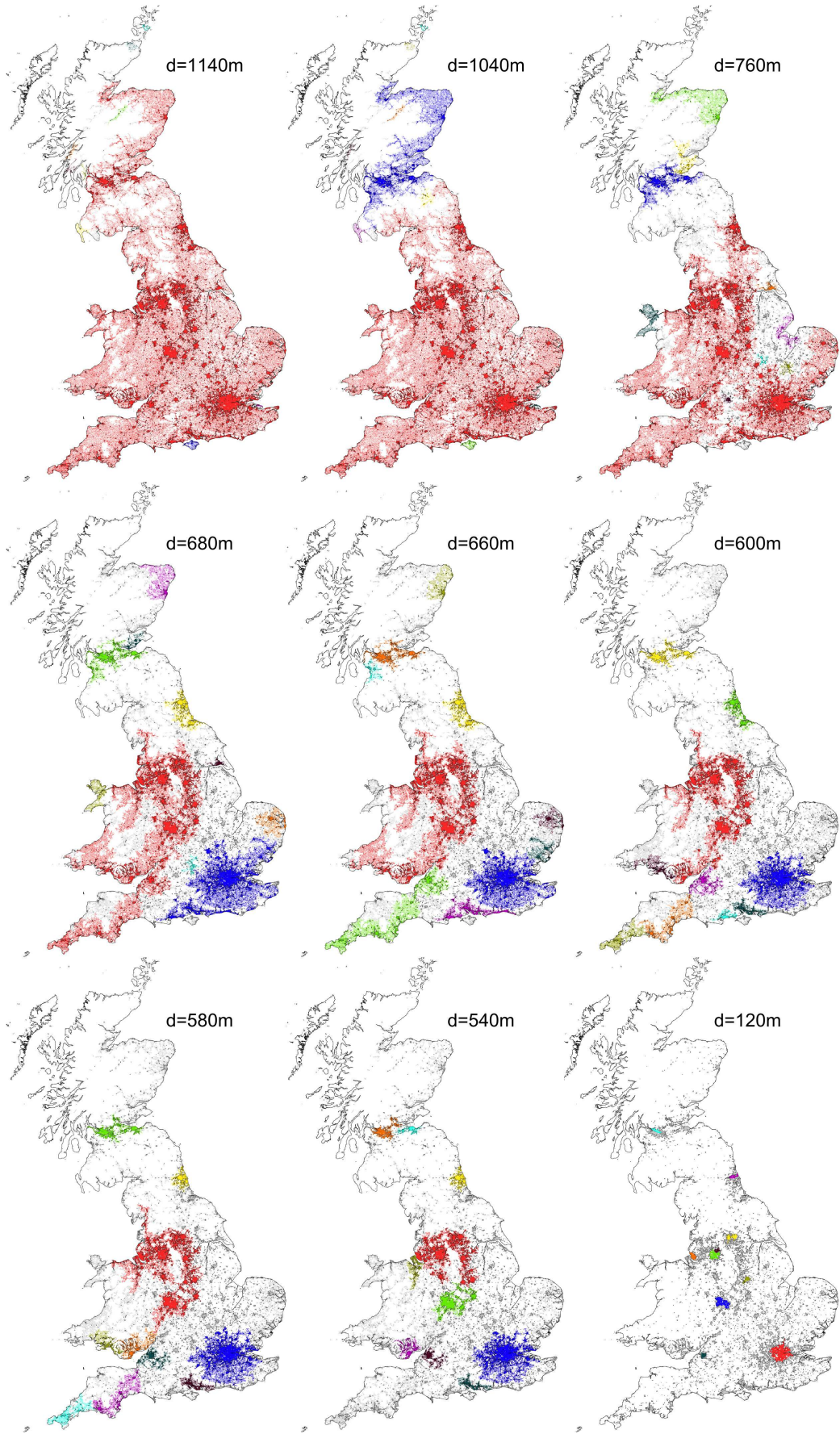}
\caption{Maps of clusters at the transitions for the percolation on the intersection points. Only the 10 largest clusters have colours following the legend in the hierarchical tree.}\label{MapsGridPerc}
\end{figure}

As discussed in the main text, although the critical distances differ for both systems (the percolation on the intersection points and the percolation on the network), see Fig.~\ref{Smax_both}, the results are very similar. The maps in Fig.~\ref{MapsGridPerc} and Fig.~\ref{MapsNetPerc} represent the transitions for the percolation on the intersection points and on the network respectively.

\begin{figure}[t]
\centering
	\includegraphics[height=0.95\textheight]{mapsTransNet.eps}
\caption{Maps of clusters at transitions for the network percolation. Only the 10 largest clusters have colours following the legend in the hierarchical tree.}\label{MapsNetPerc}
\end{figure} 

\subsection{Correlation measure of the network percolation clusters and the urban area}\label{CorrCorine}

For a given percolation result on the network, we categorise the type of each intersection as either being urban or non-urban. We define an urban intersection as an intersection that belongs to a cluster that is larger than Smin = 50 while the rest of the intersections are considered to be non-urban.

\begin{figure}[ht]
\centering
	\includegraphics[width=1\linewidth]{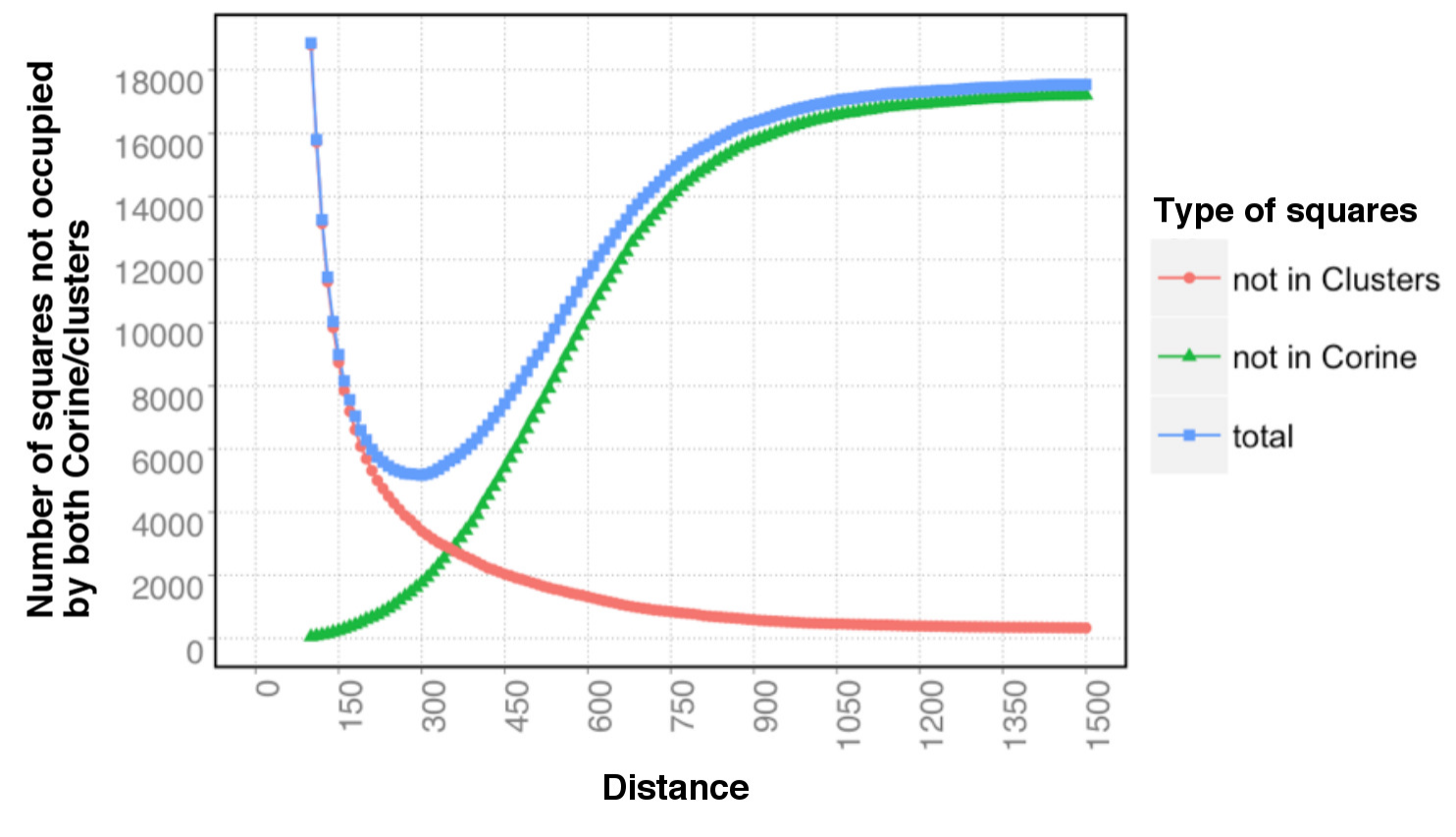}
\caption{Measure of error of concurrency between the network percolation clusters and the urbanised clusters according to the Corine dataset.}\label{errCorineClusters}
\end{figure}

We use the polygons defined in the Corine dataset as a reference point. We generate a grid of 1km per 1km over the whole territory of the UK. For each square of the grid we assign two values, the first value is the area of the polygon that corresponds to the Corine cluster that intersects the square of the grid, the second value is the mass (the number of intersections) of the percolation cluster that has more intersections in the square of the grid. In order to be able to compare both systems we perform two types of analysis. The first is a Pearson Product Moment correlation between the values assigned to each square of the grid when there is both a cluster from the Corine and a percolation cluster. 
Fig.~\ref{R2_Corine_NTS} shows that the highest correlation, for $R^2 > 0.7$ is given for the range of distances 300m to 400m. 
The second type is a measure of error comparing both. The procedure is as follows: when the squares of the grid do not have both types of clusters, we count the number of squares that have a Corine cluster but not a percolation cluster; and also the other way round, we count the number of squares that have a percolation cluster but not a Corine cluster. Finally we add both counts to get the total number of squares that do not have coincident clusters. The result is given in Fig.~\ref{errCorineClusters}, and this shows that the total number of non-coincident clusters is minimised for $d=300$m.

\subsection{Average cluster size and cluster distribution}

A traditional approach to detect transitions in percolation processes, is to look at the average cluster size removing the largest component. In order to avoid very small clusters that hold no information with respect to the hierarchical structure of the urban system under consideration, we impose a minimum cluster size. We select this minimum size to be $\text{S}_{\text{min}}=600$ intersections, since it gives enough resolution, and includes very small settlements. To put this number into context, the number of intersection points in large cities is of the order of $10^5$ and of $10^4$ for the 30 largest ones.

\begin{figure}[ht]
\includegraphics[width=0.5\linewidth]{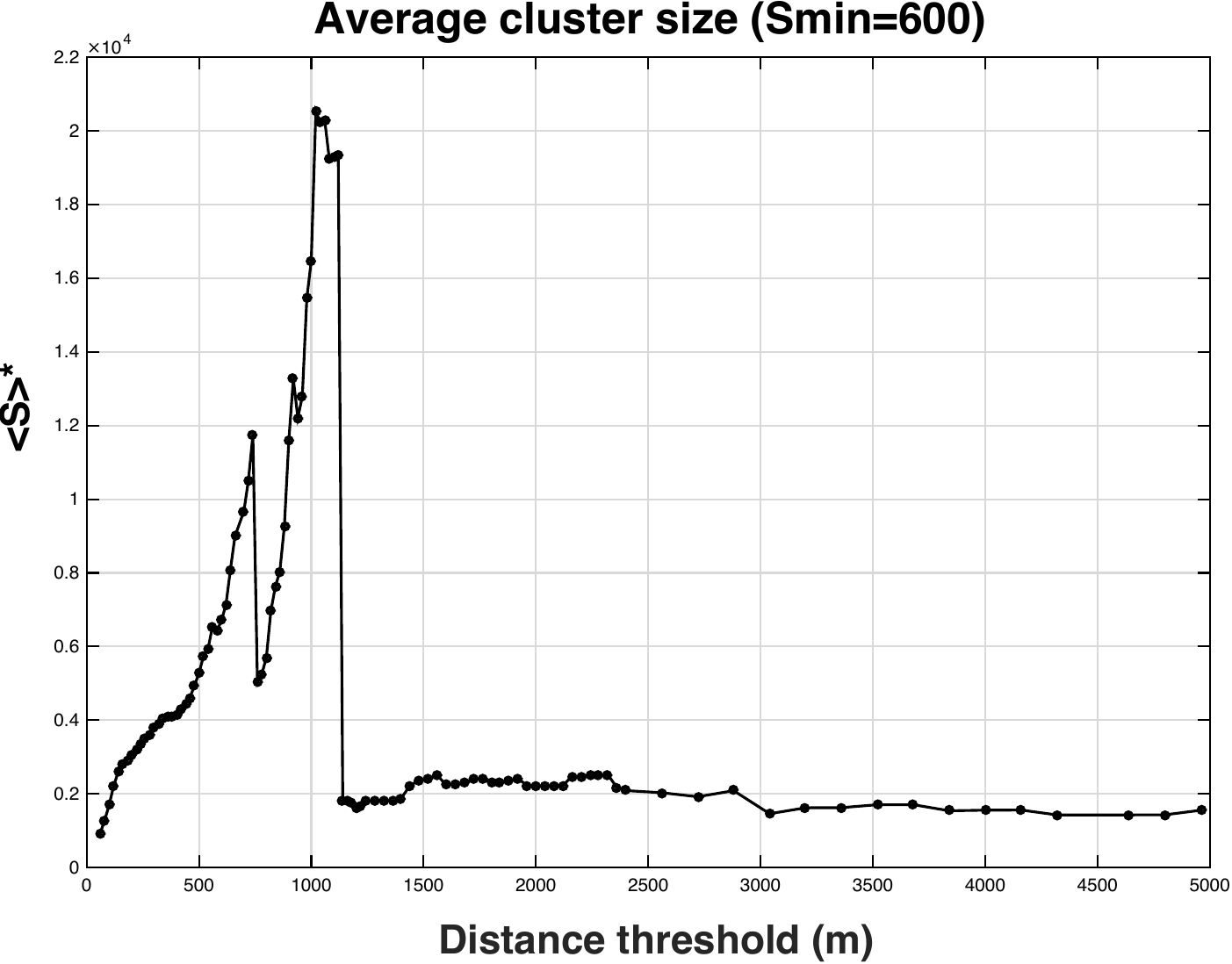}
\includegraphics[width=0.5\linewidth]{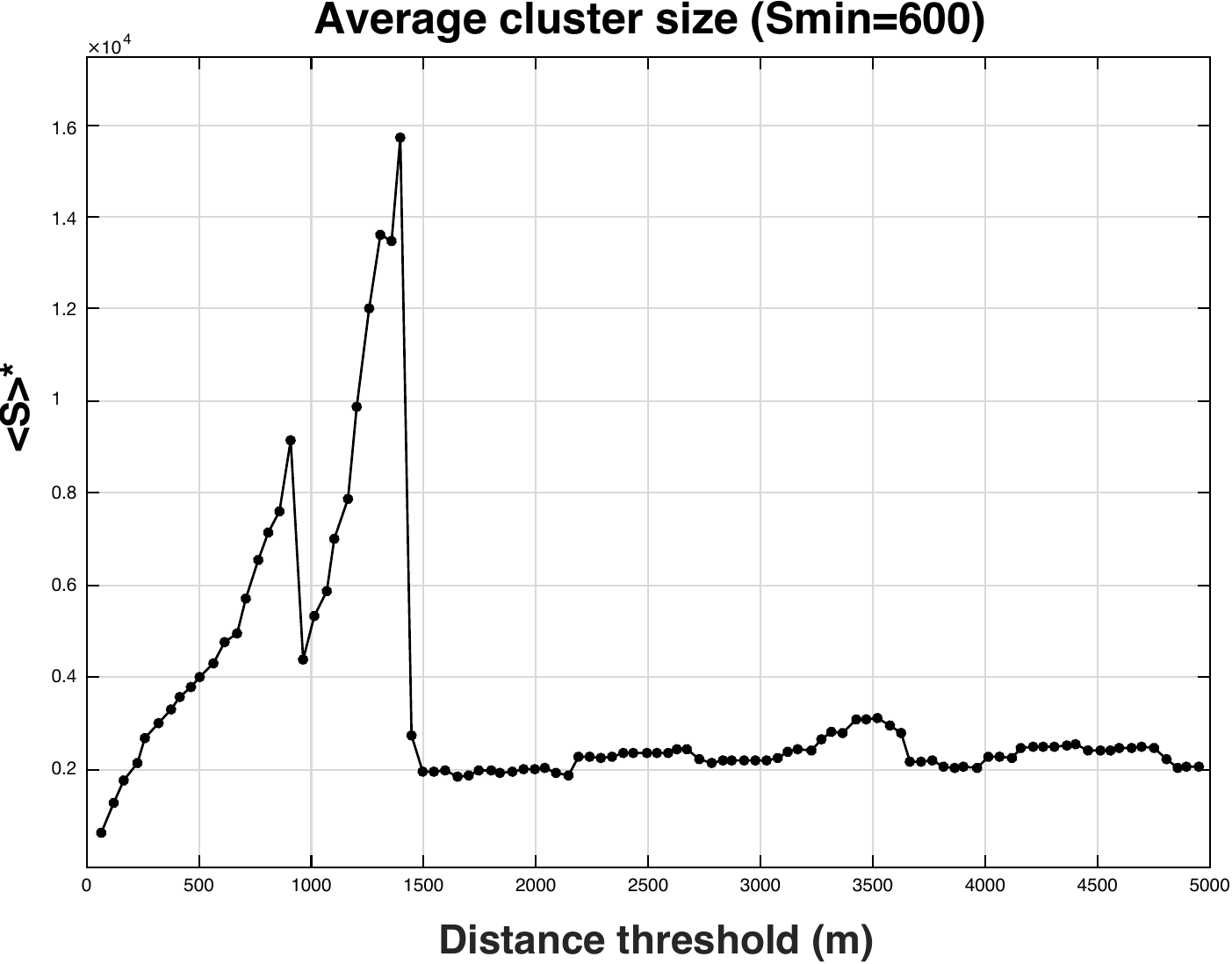}
\caption{Evolution of the average cluster size removing the largest cluster, and including only clusters with at least 600 intersections. Top, plot for the percolation on the intersection points; bottom, plot for the percolation on the network.}\label{avClusterSize}
\end{figure}

The results for both methods are given in Fig.~\ref{avClusterSize}.
Given the multiplicity of transitions arising from the different merging processes, the curve presents many different peaks. The different sizes of the second largest clusters after transitions take place, obscure many of the transitions that take place in the system. An overall picture becomes easier to grasp if one looks at the evolution of the largest cluster size, as done in Fig.~\ref{Smax_pts}.

\begin{figure}[ht]
\includegraphics[width=1\linewidth]{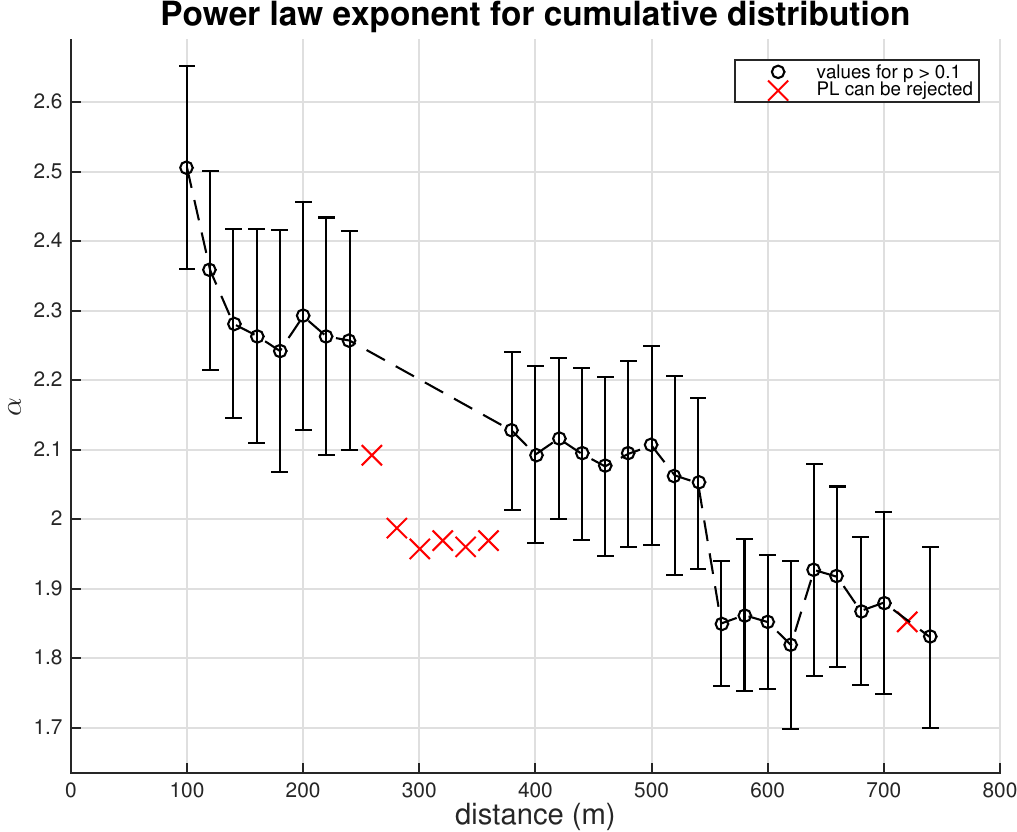}
\caption{Values with confidence intervals for the power law exponent of the cumulative distribution using the method by Clauset {\it{et al}} in \cite{Clauset_etalPL09}. If $p\leq 0.1$ the power law can be rejected, and these are the red crosses.}\label{alphaPL_d}
\end{figure}

\begin{figure}[ht]
\centering
\includegraphics[width=0.42\linewidth]{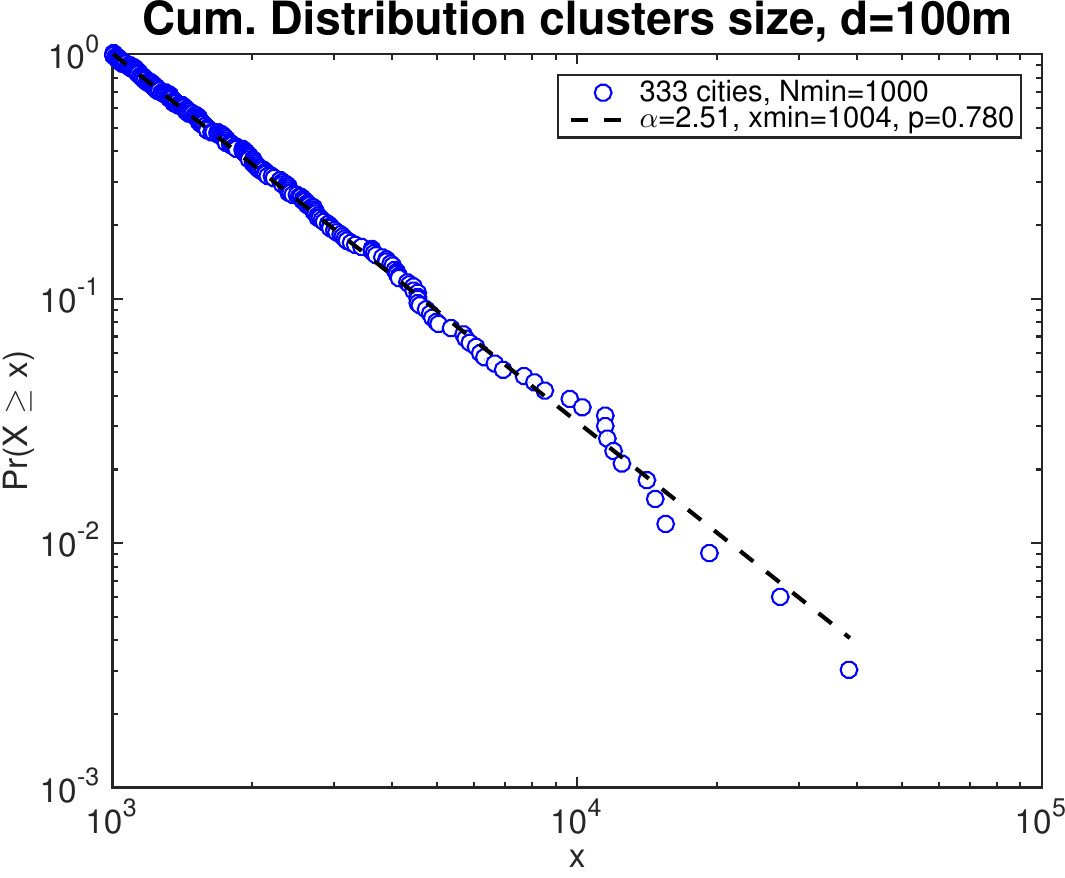}
\includegraphics[width=0.42\linewidth]{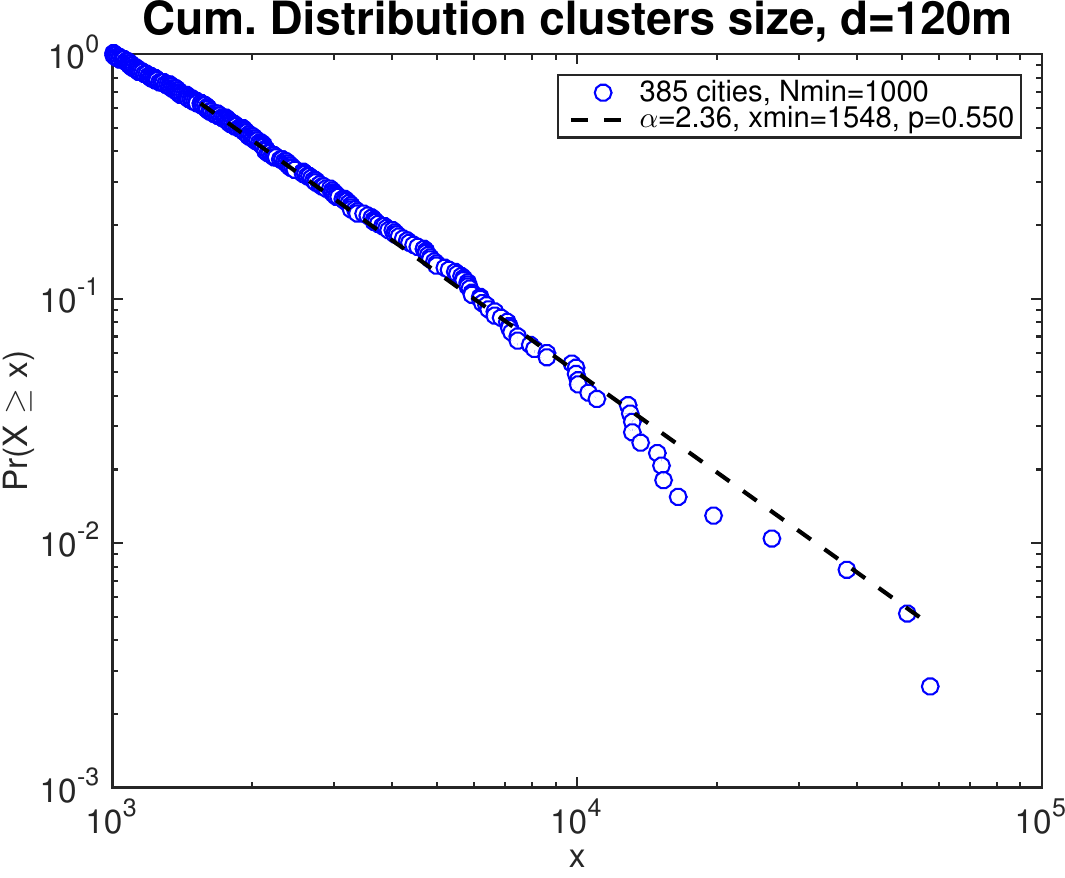}
\includegraphics[width=0.42\linewidth]{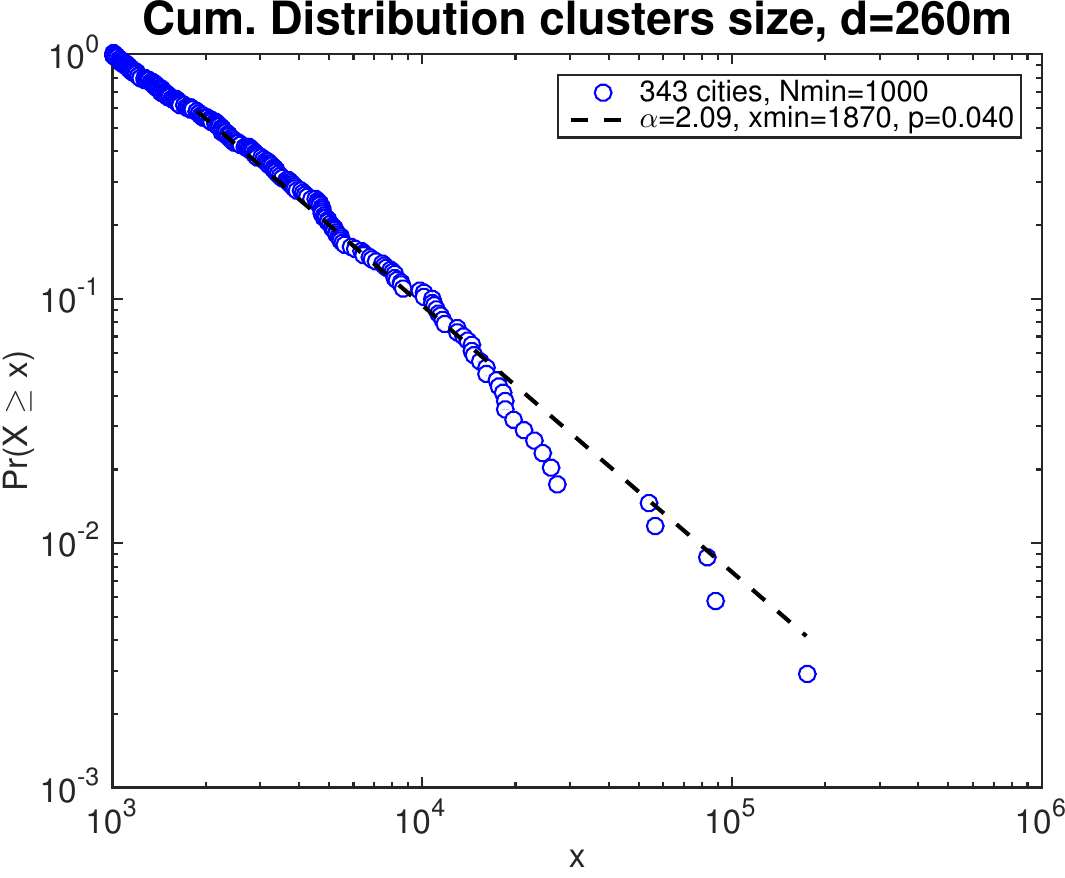}
\includegraphics[width=0.42\linewidth]{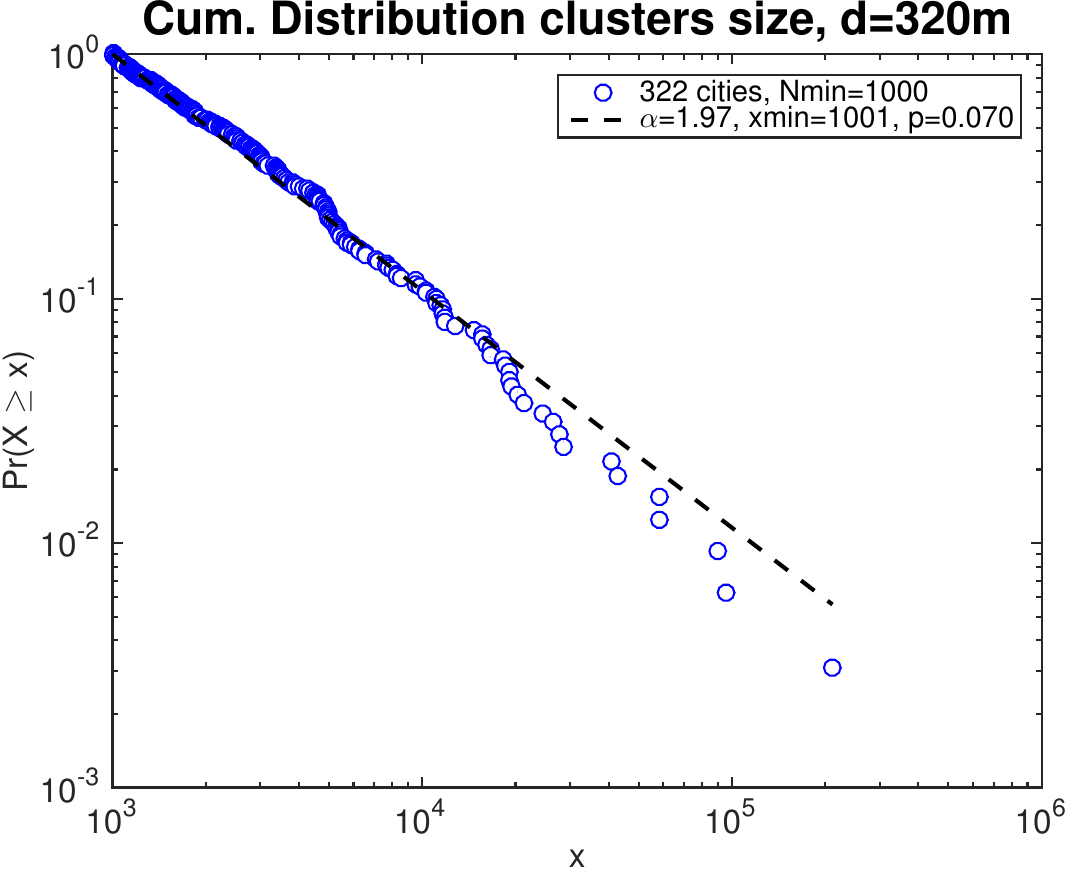}
\includegraphics[width=0.42\linewidth]{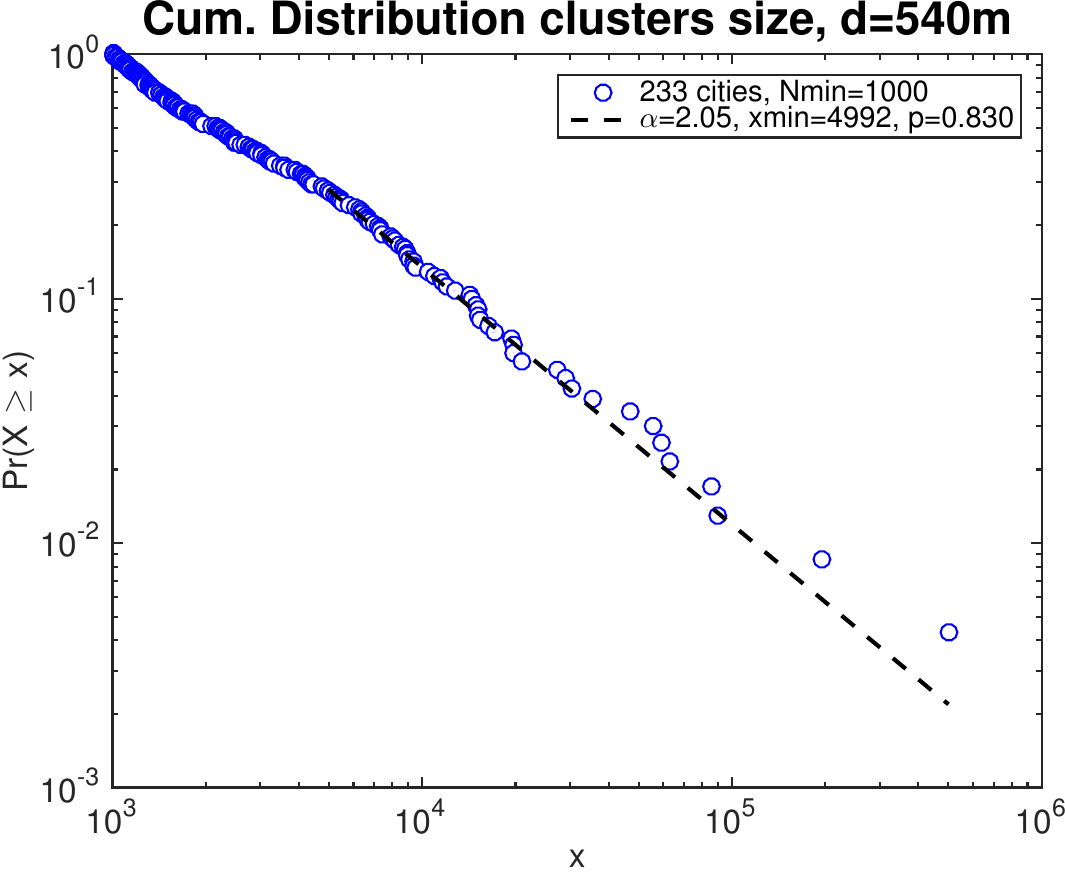}
\includegraphics[width=0.42\linewidth]{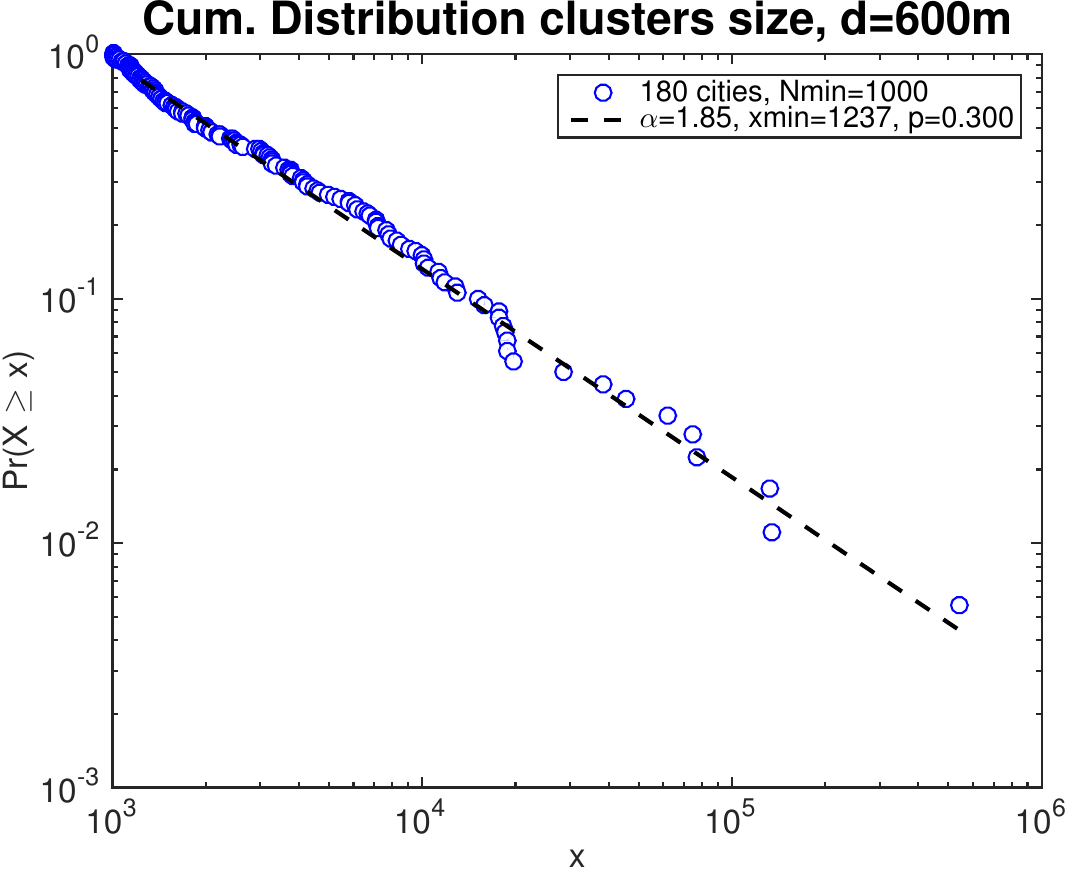}
\includegraphics[width=0.42\linewidth]{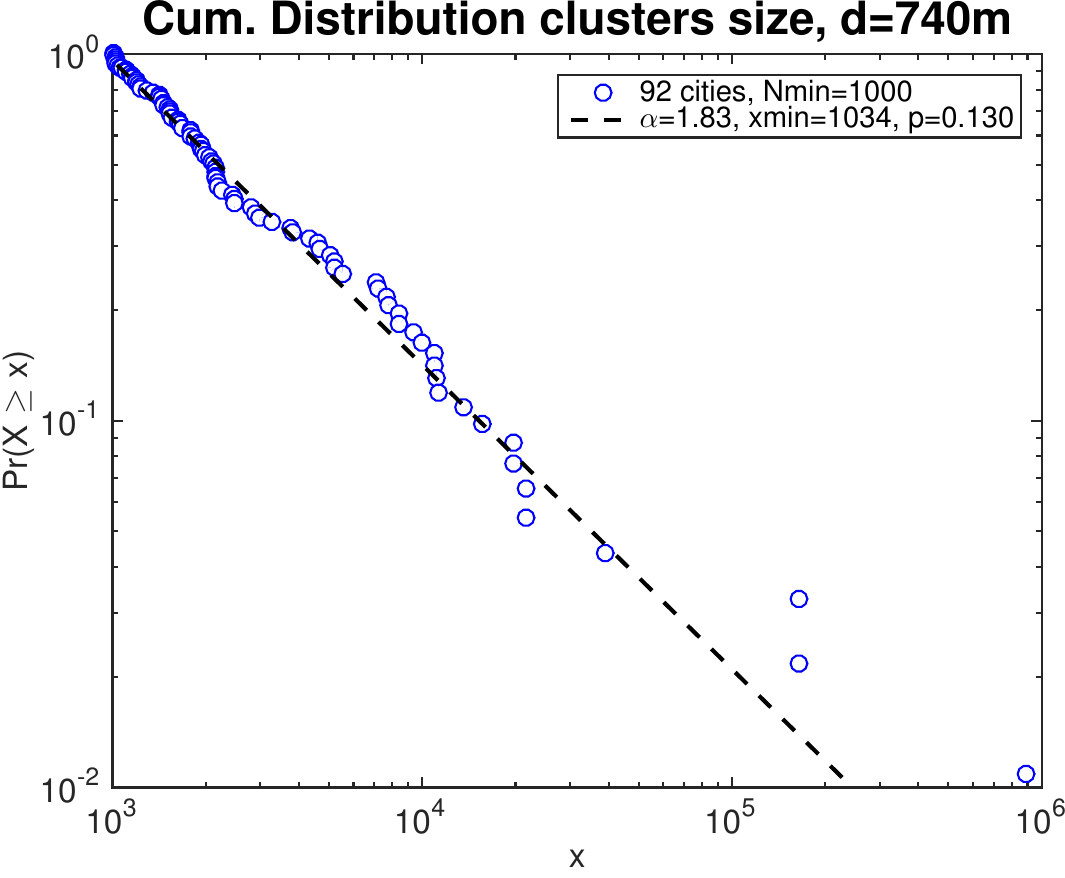}
\includegraphics[width=0.42\linewidth]{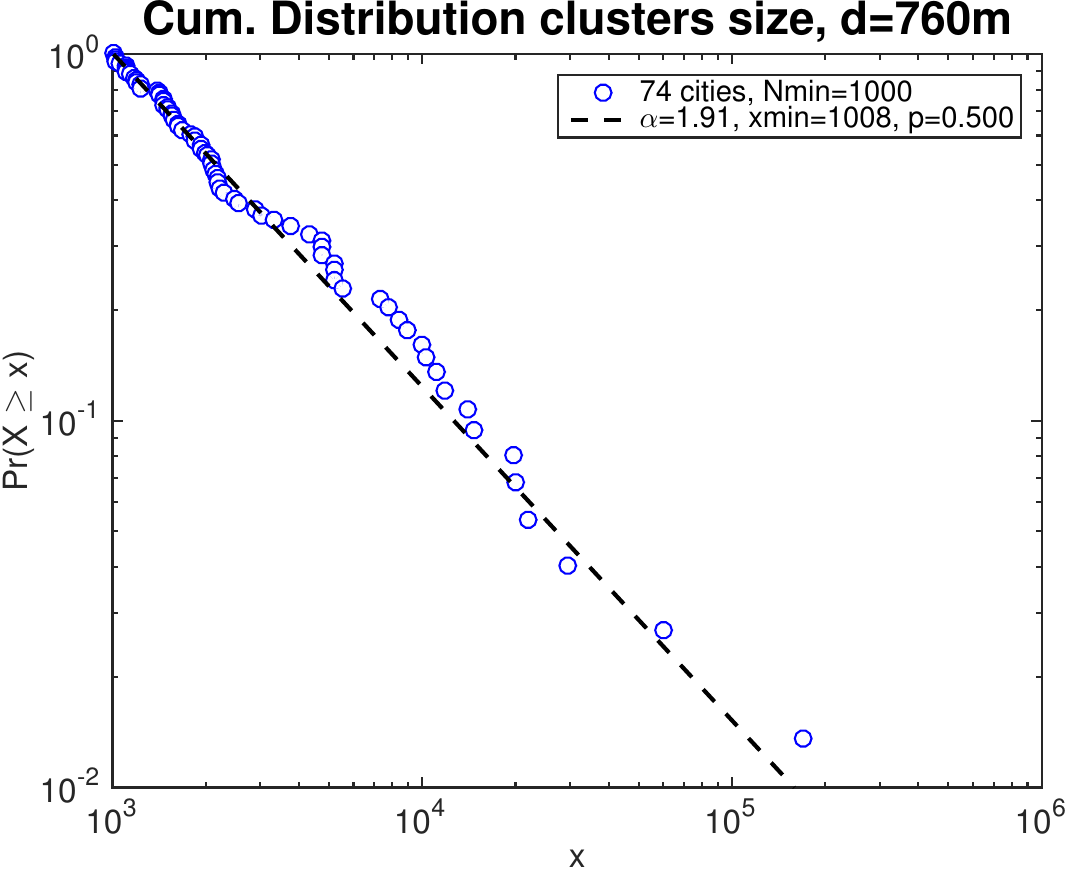}
\caption{Cumulative distribution of the clusters size removing the largest cluster, and using the Clauset {\it{et al}} method \cite{Clauset_etalPL09}.}\label{PLfit_figs}
\end{figure}

Let us now look at the distribution of cluster sizes. We investigate whether these are power law distributed. We use the method developed by \cite{Clauset_etalPL09}, where a power law can be ruled out if $p\leq 0.1$.
Note that $p> 0.1$ does not guarantee that the distribution follows a power law.
We compute the distribution for clusters that have at least 1000 points, and we remove in all cases the largest cluster, so that the giant cluster is never considered.
Given that we have a multiplicity of transitions, the second largest, and sometimes the top largest ones, can still be quite large compared to the rest of the clusters, especially for large distances. 
The results, up to distance $d=760m$ are presented in Fig.~\ref{alphaPL_d}. 
The cumulative distributions for some of the distances are given in Fig.~\ref{PLfit_figs}.
At the transitions, we note that a power law cannot be rejected.
For small distances, we observe that around the transition, the cluster sizes are power law distributed. 
There is a clear region of distances after cities formed and the next transition occurs, at which the sizes are not distributed as a power law. 
The merging mechanism leading to the multiplicity of percolation transitions translates into a fluctuating exponent of the system. It is important to remember, that exponents arising from power laws are always very sensitive to the sample considered in the distribution, the number of events etc. In this case, we only consider clusters that have at least 1000 points and we always remove the largest one. The number of clusters hence vary enormously from distance to distance. In addition, one could argue that given the multiplicity of transitions, removing the largest cluster still leaves us with very large remaining ones that will be responsible for the next transitions; this is most evident for large $d$s. 

In conclusion, the value of the exponent does not define the threshold at which cities are defined, nor whether the system is composed of cities or regions. It is important to recall that an urban system obeying a perfect Zipf law will have an exponent around 2 for the cumulative distribution of population size. 
The observed fluctuations hence contribute to the  debate of whether cities are universally distributed according to a Zipf law or not, and certainly tell us that an urban system distributed according to a Zipf law does not necessarily represent cities. 
One might argue that cities distributed according to Zipf's law is a necessary but not a sufficient condition.

\subsection{Multifractal spectrum, dimensions: $D_0$, $D_1$ and $D_2$}\label{MultFractalReview}

In this section we provide a summary of the main mathematical expressions to compute the fractal dimensions $D_0$, $D_1$ and $D_2$. This is an extract from \cite{Murcio_Multifractal2015}, where we give a detailed review of the methodology employed in order to compute these three measures from the whole multifractal spectrum. In summary, a (mono-) fractal has a measure that is homogeneous in space, such that within a small region $\epsilon$
\begin{equation}
\mu (\epsilon) \sim \epsilon^{-D}
\end{equation}
where $D$ is the fractal dimension. For a multifractal, $\mu$ is no longer homogeneous, and hence for each region $i$ we can define a distribution function $P_i(\epsilon)$ of the measure
\begin{equation}
P_i(\epsilon) \sim \epsilon^{\alpha_i}
\end{equation}
where each subdivision of the space $i$ has a value $\alpha_i$. A fractal dimension $f(\alpha_i)$ can then be associated for the set of regions with the same value. The moments of the distribution function are obtained through the function
\begin{equation}
Z_q(\epsilon)=\sum_i P_i(\epsilon)^q \sim (\epsilon)^{-\tau(q)}
\end{equation}
The exponent $\tau(q)$ can be written in terms of the generalised fractal dimension $D_q$ as
\begin{equation}
\tau(q)=q\alpha(q)-f[\alpha(q)]=(q-1)D_q
\end{equation}
where
\begin{equation}
D_q=\frac{1}{q-1}\lim_{\epsilon \rightarrow 0} \frac{\log_{10}[Z_q(\epsilon)]}{\log_{10}(\epsilon)}
\end{equation}
defines the whole spectrum for $q \in (-\infty, \infty)$. The 3 fractal dimensions are hence obtained for $q \in \{0, 1, 2\}$. For a monofractal, $D_q$ is a constant for all $q$s. For $q=0$, the fractal dimension $D_0$ corresponds to the dimension obtained through a box-counting algorithm. For $q=1$, we get
\begin{equation}
D_1=\lim_{\epsilon \rightarrow 0}\frac{-\sum_iP_i\log_{10}P_i}{-\log_{10}(\epsilon)}
\end{equation}
which has a very similar form to Shannon's entropy, and this is why this is called the information dimension. Finally, for $q=2$, the dimension $D_2$ takes the form
\begin{equation}
D_2=\lim_{\epsilon \rightarrow 0}\frac{\sum_i P_i^2}{\log_{10}(\epsilon)}
\end{equation}
and this in our case gives the correlations for pairs of intersection points to lie within the same box $\epsilon$.

\end{document}